\documentstyle[psfig]{mn}

\newcommand{\bvvmax}{$\langle (V-V_0)/(V_{\rm max}-V_0) \rangle$} 
\newcommand{\chisq}{$\chi^2$}			
\newcommand{\etal}{et~al.}			
\newcommand{\flilly}{$f_{5000}$}		
\newcommand{\fourang}{4000-\AA\ break}		
\newcommand{\fsh}{4-Shooter}			
\newcommand{\kmsmpc}{km s$^{-1}$ Mpc$^{-1}$}	
\newcommand{\logten}{\mathop{{\rm log}_{10}}}	
\newcommand{\Lya}{\hbox{Ly$\alpha$}}		
\newcommand{\micron}{$\mu$m}			
\newcommand{\rlf}{$\rho(P,z)$}			
\newcommand{\vvmax}{$V/V_{\rm max}$}		
\newcommand{\WHzsr}{W~Hz$^{-1}$~sr$^{-1}$}	

\newcommand{\ione}{$\,${\sc i}}			
\newcommand{\ii}{$\,${\sc ii}}
\newcommand{\iii}{$\,${\sc iii}}
\newcommand{\iv}{$\,${\sc iv}}
\newcommand{\vv}{$\,${\sc v}}

\newcommand{\aap}{A\&A}
\newcommand{\aapss}{A\&A, Suppl.\ Ser.}
\newcommand{\aj}{AJ}
\newcommand{\apj}{ApJ}
\newcommand{\apjl}{ApJL}

\newcommand{\araa}{ARA\&A}
\newcommand{\mnras}{MNRAS}
\newcommand{\nat}{Nature}
\newcommand{\pasp}{PASP}

\newif\ifAMStwofonts

\ifoldfss
  \ifCUPmtlplainloaded \else
    \NewTextAlphabet{textbfit} {cmbxti10} {}
    \NewTextAlphabet{textbfss} {cmssbx10} {}
    \NewMathAlphabet{mathbfit} {cmbxti10} {} 
    \NewMathAlphabet{mathbfss} {cmssbx10} {} 
  \fi
  \ifAMStwofonts
    \ifCUPmtlplainloaded \else
      \NewSymbolFont{upmath} {eurm10}
      \NewSymbolFont{AMSa} {msam10}
      \NewMathSymbol{\upi}     {0}{upmath}{19}
      \NewMathSymbol{\umu}     {0}{upmath}{16}
      \NewMathSymbol{\upartial}{0}{upmath}{40}
      \NewMathSymbol{\leqslant}{3}{AMSa}{36}
      \NewMathSymbol{\geqslant}{3}{AMSa}{3E}

       \let\le=\leqslant
       \let\ge=\geqslant
    \fi
  \fi
\fi 

\ifnfssone
  \newmathalphabet{\mathit}
  \addtoversion{normal}{\mathit}{cmr}{m}{it}
  \addtoversion{bold}{\mathit}{cmr}{bx}{it}
  \newmathalphabet{\mathbfit} 
  \addtoversion{normal}{\mathbfit}{cmr}{bx}{it}
  \addtoversion{bold}{\mathbfit}{cmr}{bx}{it}
  \newmathalphabet{\mathbfss} 
  \addtoversion{normal}{\mathbfss}{cmss}{bx}{n}
  \addtoversion{bold}{\mathbfss}{cmss}{bx}{n}
  \ifAMStwofonts
    \ifCUPmtlplainloaded \else
      \UseAMStwoboldmath
      \makeatletter
      \new@mathgroup\upmath@group
      \define@mathgroup\mv@normal\upmath@group{eur}{m}{n}
      \define@mathgroup\mv@bold\upmath@group{eur}{b}{n}
      \edef\UPM{\hexnumber\upmath@group}
      \new@mathgroup\amsa@group
      \define@mathgroup\mv@normal\amsa@group{msa}{m}{n}
      \define@mathgroup\mv@bold\amsa@group{msa}{m}{n}
      \edef\AMSa{\hexnumber\amsa@group}
      \makeatother
      \mathchardef\upi="0\UPM19
      \mathchardef\umu="0\UPM16
      \mathchardef\upartial="0\UPM40
      \mathchardef\leqslant="3\AMSa36
      \mathchardef\geqslant="3\AMSa3E

       \let\le=\leqslant
       \let\ge=\geqslant
    \fi
  \fi
\fi 

\ifnfsstwo
  \DeclareMathAlphabet{\mathbfit}{OT1}{cmr}{bx}{it}
  \SetMathAlphabet\mathbfit{bold}{OT1}{cmr}{bx}{it}
  \DeclareMathAlphabet{\mathbfss}{OT1}{cmss}{bx}{n}
  \SetMathAlphabet\mathbfss{bold}{OT1}{cmss}{bx}{n}
  \ifAMStwofonts
    \ifCUPmtlplainloaded \else
      \DeclareSymbolFont{UPM}{U}{eur}{m}{n}
      \SetSymbolFont{UPM}{bold}{U}{eur}{b}{n}
      \DeclareSymbolFont{AMSa}{U}{msa}{m}{n}
      \DeclareMathSymbol{\upi}{0}{UPM}{"19}
      \DeclareMathSymbol{\umu}{0}{UPM}{"16}
      \DeclareMathSymbol{\upartial}{0}{UPM}{"40}
      \DeclareMathSymbol{\leqslant}{3}{AMSa}{"36}
      \DeclareMathSymbol{\geqslant}{3}{AMSa}{"3E}

       \let\le=\leqslant
       \let\ge=\geqslant
    \fi
  \fi
\fi 

\ifCUPmtlplainloaded \else
  \ifAMStwofonts \else 
    \def\upi{\pi}
    \def\umu{\mu}
    \def\upartial{\partial}
  \fi
\fi

\title[LBDS Hercules -- II. Redshift distribution]{The LBDS Hercules
sample of mJy radio sources at 1.4~GHz -- II. Redshift distribution,
radio luminosity function, and the high-redshift cut-off}

\author[Waddington et al.]{I. Waddington$^{1,2}$\thanks{Present
address: Department of Physics, University of Bristol, H. H. Wills
Physics Laboratory, Tyndall Avenue, Bristol BS8 1TL, UK.  Email:
I.Waddington@bristol.ac.uk}, J. S. Dunlop$^2$, J. A. Peacock$^2$, and
R. A. Windhorst$^1$\\ $^1$ Department of Physics \& Astronomy, Arizona
State University, PO Box 871504, Tempe, AZ 85287--1504, USA\\ $^2$
Institute for Astronomy, University of Edinburgh, Royal Observatory,
Blackford Hill, Edinburgh EH9 3HJ, UK}

\date{Accepted - .
      Received - ;
      in original form -}

\pagerange{\pageref{firstpage}--\pageref{lastpage}}
\pubyear{2001}

\begin{document}

\maketitle

\label{firstpage}

\begin{abstract}
A combination of spectroscopy and broadband photometric redshifts has
been used to find the complete redshift distribution of the Hercules
sample of millijansky radio sources.  These data have been used to
examine the evolution of the radio luminosity function (RLF) and its
high-redshift cut-off.

We report the results of recent spectroscopic observations of the
Hercules sample, drawn from the 1.4~GHz Leiden-Berkeley Deep Survey
(LBDS) with a flux-density limit of 1~mJy.  New redshifts have been
measured for eleven sources, and a further ten were detected in
continuum emission from which upper limits to the redshift are given,
derived from the absence of a Lyman-limit break in their spectra.  The
total number of sources with known redshifts in the sample is now 47
(65\%).  We calculated broadband photometric redshifts for the
remaining one-third of the sample, using a two-component (old stellar
population plus starburst) spectral synthesis model.

We use the resulting redshift distribution of this complete sample to
investigate the cosmological evolution of radio sources.  For the
luminosity range probed by the present study ($P_{\rm 1.4~ GHz} >
10^{24.5}$~\WHzsr), we use the $V/V_{max}$ test to show conclusively
that there is a deficit of high-redshift ($z > 2$--2.5) objects.

Comparison with the model RLFs of Dunlop \& Peacock (1990) shows that
our data can now exclude pure luminosity evolution as a viable
description of the cosmological evolution of the RLF.  However two of
the models of DP90 successfully predict the redshift-dependent
evolution of the millijansky population and are approximately
consistent with its observed luminosity-dependence.  These models, and
the RLF deduced by direct binning of the data, both favour a
luminosity dependence for the high-redshift cut-off, with
lower-luminosity sources ($P_{\rm 1.4~GHz} \simeq 10^{24}$~\WHzsr) in
decline by $z \simeq 1$--1.5 while higher-luminosity sources ($P_{\rm
1.4~GHz} \simeq 10^{25-26}$~\WHzsr) decline in comoving number density
beyond $z \simeq 2$--2.5.
\end{abstract}


\section{Introduction}

It has long been known that the comoving number density of radio
sources increases with redshift.  Evidence for a subsequent decline in
the comoving density of low luminosity radio galaxies at $z\ga 1$ was
presented by Windhorst (1984)\nocite{Windhorst84}.  At high
luminosities, Peacock (1985)\nocite{Peacock85} concluded that there
was a reduction in the comoving density of compact, flat-spectrum
($\alpha<0.5$, where $S_\nu \propto \nu^{-\alpha}$), radio-loud
quasars at $z\ga 2$.  Dunlop \& Peacock (1990, hereafter
DP90)\nocite{Dunlop90} extended that study to lower radio luminosities
and steep-spectrum sources, and presented evidence that there is also
a peak at $z\simeq 2$ in the density of both steep-spectrum quasars
and radio galaxies.

However the reality of the decline or `cut-off' in the radio
luminosity function (RLF) at high redshift (which {\it cannot\/} be
due to dust obscuration, for example) has been challenged following
the discovery of radio galaxies at $z>4$ (e.g., Spinrad, Dey, \&
Graham 1995; van Breugel \etal\ 1999).\nocite{Spinrad95,vanBreugel99}
Although limited by small-number statistics, Jarvis
et~al.~(2001)\nocite{Jarvis01} find that the comoving number density
of the most powerful ($P_{\rm 1.4~GHz} > 10^{27}$~\WHzsr)
steep-spectrum sources remains constant between $z\sim 2.5$ and $z\sim
4.5$.

The best way to settle this uncertainty is to determine the redshift
distribution of a complete sample of radio galaxies selected at a
flux-density limit of $\sim 1$~mJy.  This limit is sufficiently {\it
faint\/} that, if there is no redshift cut-off, a large fraction of
the sources must lie at high redshifts, and yet sufficiently {\it
bright\/} that objects selected at this flux density are still
predominantly classical radio galaxies rather than low-redshift
starbursts \cite{Windhorst85,Hopkins00}.

The Leiden-Berkeley Deep Survey (LBDS) consists of nine high-latitude
fields in the selected areas SA28, SA57, SA68 and an area in Hercules.
They were surveyed with the Westerbork Synthesis Radio Telescope at
1.41~GHz, reaching a 5-$\sigma$ limiting flux density of 1~mJy
(Windhorst, van Heerde \& Katgert 1984a)\nocite{Windhorst84a}.
Optical counterparts to the radio sources were found using
multi-colour prime focus photographic plates of the fields, taken with
the Kitt Peak 4-m Mayall telescope.  Identifications were found for
53\% of the sources in the full survey, whilst for the Hercules field
47 out of 72 sources were identified (Windhorst, Kron \& Koo 1984b;
Kron, Koo \& Windhorst 1985)\nocite{Windhorst84b,Kron85}.

The Hercules field was subsequently observed with the 200-inch Hale
Telescope at Palomar Observatory between 1984 and 1988 (Waddington
\etal\ 2000, hereafter Paper~I)\nocite{Waddington00}.  Multiple
observations were made through Gunn $g$, $r$ and $i$ filters over six
observing runs.  After processing and stacking of the multiple-epoch
images, optical counterparts for 22 of the sources were found, leaving
only three sources unidentified to $r\simeq 26$~mag.  Near-infrared
observations have been made of the entire Hercules sample at $K$,
yielding 60 (of 72) detections down to $K\simeq 19$--21~mag.  Half of
the sources have been observed in $H$ and approximately one-third in
$J$ (Paper~I).  Prior to the work reported here, approximately half of
the sample had spectroscopic redshifts, which were summarised in
Table~4 of Paper~I.  In this paper, we present additional redshift
measurements together with photometric redshift estimates, and then
investigate the redshift distribution of the sample.

The layout of the paper is as follows.  In section~2 we present the
results of our most recent efforts to obtain optical spectra of the
radio sources with the William Herschel Telescope.  In section~3 we
describe the methods that we used to calculate photometric redshifts
for the 22 sources for which we do not have a spectroscopic
measurement.  The complete redshift distribution (spectroscopic plus
photometric) is then discussed.  Finally, in section~4 we compare the
redshift distribution of this millijansky sample with the predictions
of the model radio luminosity functions of Dunlop \& Peacock (1990),
and reconsider the evidence for a redshift cut-off in the RLF.  For
consistency with earlier papers, unless otherwise stated, a cosmology
with $H_0 = 50$~\kmsmpc, $\Omega_0 = 1$ and $\Omega_\Lambda = 0$ is
assumed.

\section{Spectroscopic data}

\subsection{Observations and data reduction}

Observations of 7 faint red radio galaxies in the LBDS Hercules sample
were obtained with the ISIS spectrograph at the 4.2-m William Herschel
Telescope (WHT) at the Observatorio del Roque de Los Muchachos, La
Palma, on 1995 June 21--24.  On 1997 June 30 -- July 3, a further 22
sources were observed with ISIS at the WHT, using the same
configuration.  ISIS is a double-beam spectrograph -- light from the
slit is split by a dichroic at 6100~\AA, with the blue and red light
going into separate channels where the optical coatings and CCD
detectors have been sensitised to blue and red wavelengths
respectively.  Using the R158B \& R158R gratings, the total wavelength
coverage was 3160--9060~\AA\ with a dispersion of
2.9~\AA~pixel$^{-1}$.

Each observation consisted of between one and six exposures of 1800~s
each on the radio galaxy, giving a total integration time of 0.5--3
hours per target.  The spectra were initially processed online
throughout the night, enabling us to move to the next target as soon
as sufficient signal was obtained to determine a redshift.  The
targets were acquired by blind-offsetting from a nearby star, which
was also used to determine the expected position of the source on the
CCD.  Spectrophotometric standards were observed at the start of each
night.  Arc spectra, tungsten flat-fields and twilight sky flat-fields
were taken at the start and end of each night.

The data were reduced using standard techniques.  After bias
subtraction and flat-fielding, the one-dimensional spectrum was
extracted with the NOAO spectroscopy package {\sc apextract}.  An
aperture of 3~arcsec was usually sufficient to contain the whole
object, while keeping the sky to a minimum.  Sky apertures of
5--9~arcsec width were defined on each side of the object aperture.
For strong continuum sources, the object could be traced along the
full length of the dispersion axis with a low-order ($\le 3$)
polynomial, and this trace was used to extract the spectrum.  For
sources with poor continuum signal, an adjacent observation of an
offset or standard star was used to define the trace.  For the spectra
with the strongest continuum we used optimal extraction, where the
pixels were weighted by their variance, but for fainter sources the
spectrum was simply added across the spatial axis.  For the very
faintest targets (particularly some observed in the 1995 run) we used
{\sc figaro}'s {\sc polysky} application to perform a two-dimensional
sky subtraction, and then searched for line or continuum emission in
the image without extracting a one-dimensional spectrum.

Removal of the atmospheric absorption features in the red channel of
the spectrum was performed using observations of the early-type star
G138$-$31a.  Each object spectrum was calibrated by first correcting
for extinction, and then multiplying by a sensitivity function
calculated from observations of the spectrophotometric standard star
BD$+$33$\degr$2642.  Finally, the calibrated blue and red spectra were
combined by taking the mean of the 32-\AA\ overlap region.  The
calibrations were tested at this stage by applying them to the
standard star observations, and examining the quality of the joint at
6100~\AA.  There was a small discontinuity of about 8\% between the
blue and red fluxes across the joint of the standard stars.  However,
this level of photometric accuracy was quite sufficient for this
project which was concerned with measuring redshifts rather than
accurate fluxes.


\begin{table}
\caption{Redshifts and individual line identifications for the 11 LBDS
Hercules sources observed at the WHT.\label{linelist}}
\begin{tabular}{cclc} \hline
Source & $\lambda_{\rm obs}$ (\AA) & Identification & $z$ \\
Mean Redshift &  &  &  \\ \hline
53W012              & 3814 & He\ii\ 1640      & 1.326  \\
$1.328 \pm 0.001$   & 4440 & C\iii\ 1909      & 1.326  \\
                    & 5641 & Ne\iv\ 2423      & 1.328  \\
                    & 7981 & [Ne\vv] 3426     & 1.330  \\
                    & 8682 & [O\ii] 3727      & 1.330  \\
                    &      &                  &        \\
53W019              & 6056 & Ca K 3933        & 0.5400 \\
$0.5415 \pm 0.0005$ & 6117 & Ca H 3968        & 0.5416 \\
                    & 7979 & Mg b 5174        & 0.5421 \\
                    & 7722 & [O\iii] 5007     & 0.5422 \\
                    &      &                  &        \\
53W022              & 5694 & [O\ii] 3727      & 0.5278 \\
$0.5279 \pm 0.0001$ & 7428 & H$\beta$ 4861    & 0.5281 \\
                    & 7577 & [O\iii] 4959     & 0.5279 \\
                    & 7650 & [O\iii] 5007     & 0.5279 \\
                    &      &                  &        \\
53W026              & 6200 & 4000-\AA\ break  & 0.55   \\
$0.55 \pm 0.05$     &      &                  &        \\
                    &      &                  &        \\
53W027              & 4803 & [Ne\vv] 3426     & 0.4019 \\
$0.4027 \pm 0.0002$ & 5227 & [O\ii] 3727      & 0.4025 \\
                    & 6822 & H$\beta$ 4861    & 0.4034 \\
                    & 6956 & [O\iii] 4959     & 0.4027 \\
                    & 7024 & [O\iii] 5007     & 0.4028 \\
                    &      &                  &        \\
53W034              & 4775 & [O\ii] 3727      & 0.2812 \\
$0.2809 \pm 0.0001$ & 6226 & H$\beta$ 4861    & 0.2808 \\
                    & 6353 & [O\iii] 4959     & 0.2811 \\
                    & 6411 & [O\iii] 5007     & 0.2804 \\
                    & 8389 & [N\ii] 6548      & 0.2812 \\
                    & 8406 & H$\alpha$ 6563   & 0.2808 \\
                    & 8432 & [N\ii] 6583      & 0.2809 \\
                    &      &                  &        \\
53W048              & 6587 & Ca K 3933        & 0.6748 \\
$0.6755 \pm 0.0003$ & 6649 & Ca H 3968        & 0.6756 \\
                    & 7208 & G-band 4300      & 0.6763 \\
                    & 8668 & Mg b 5174        & 0.6753 \\
                    &      &                  &        \\
53W065              & 8151 & [O\ii] 3727      & 1.187  \\
$1.185 \pm 0.001$   & 6110 & Mg\ii\ 2799      & 1.183  \\
                    &      &                  &        \\
53W067              & 6551 & [O\ii] 3727      & 0.758  \\
$0.759 \pm 0.001$   & 6927 & Ca K 3933        & 0.761  \\
                    & 6983 & Ca H 3968        & 0.760  \\
                    & 7550 & G-band 4300      & 0.756  \\
                    &      &                  &        \\
53W083              & 6510 & 4000-\AA\ break  & 0.628  \\
$0.628 \pm 0.003$   &      &                  &        \\
                    &      &                  &        \\
53W089              & 6093 & [O\ii] 3727      & 0.635  \\
$0.634 \pm 0.001$   & 8097 & [O\iii] 4959     & 0.633  \\
                    & 8178 & [O\iii] 5007     & 0.633  \\ \hline
\end{tabular}
\end{table}

\subsection{Results of the spectroscopy}

\begin{figure*}
\begin{minipage}{17cm}
\hbox{\psfig{file=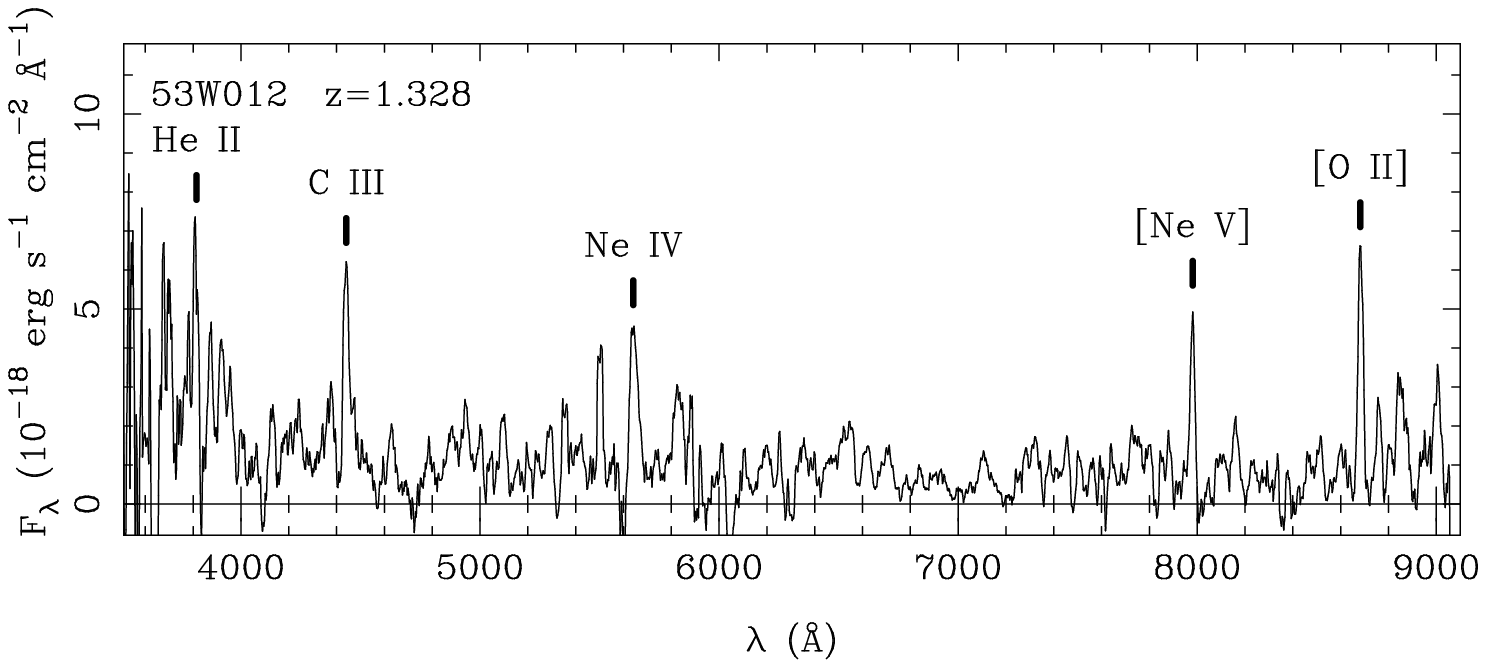,width=8.5cm}\psfig{file=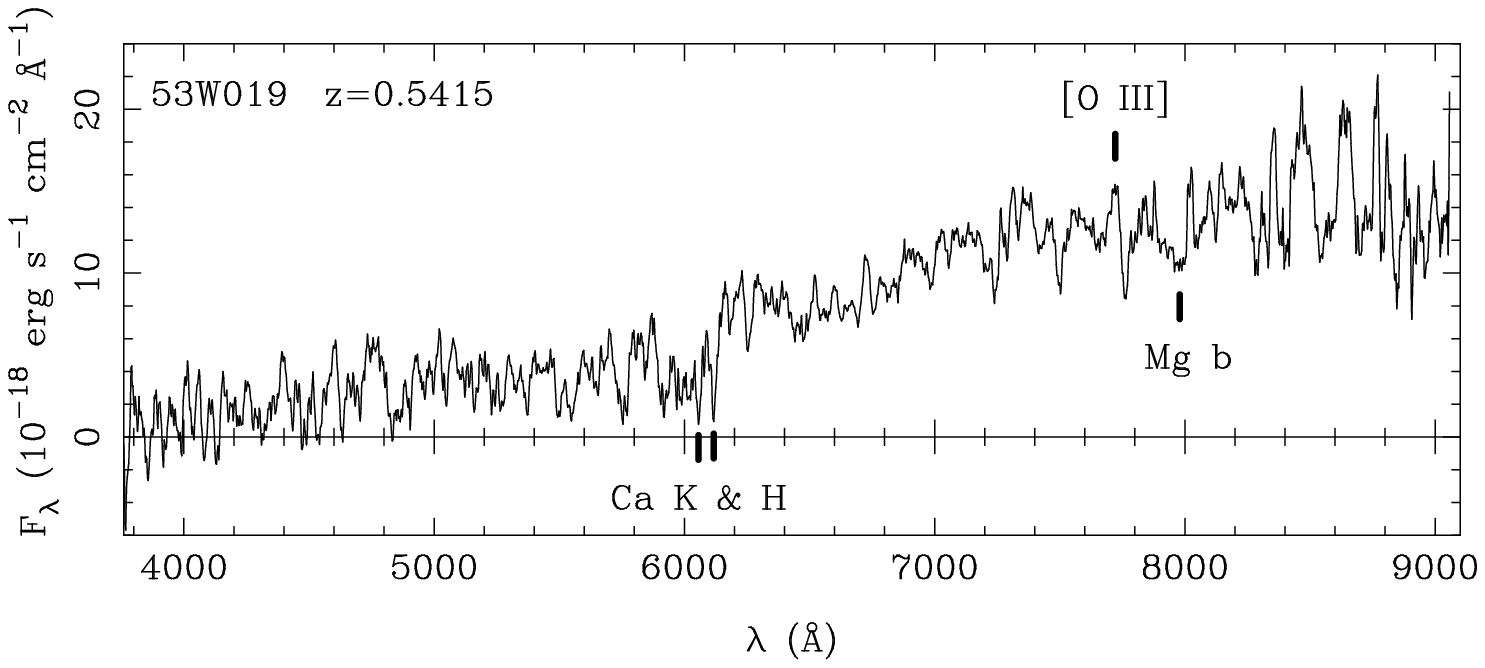,width=8.5cm}}
\hbox{\psfig{file=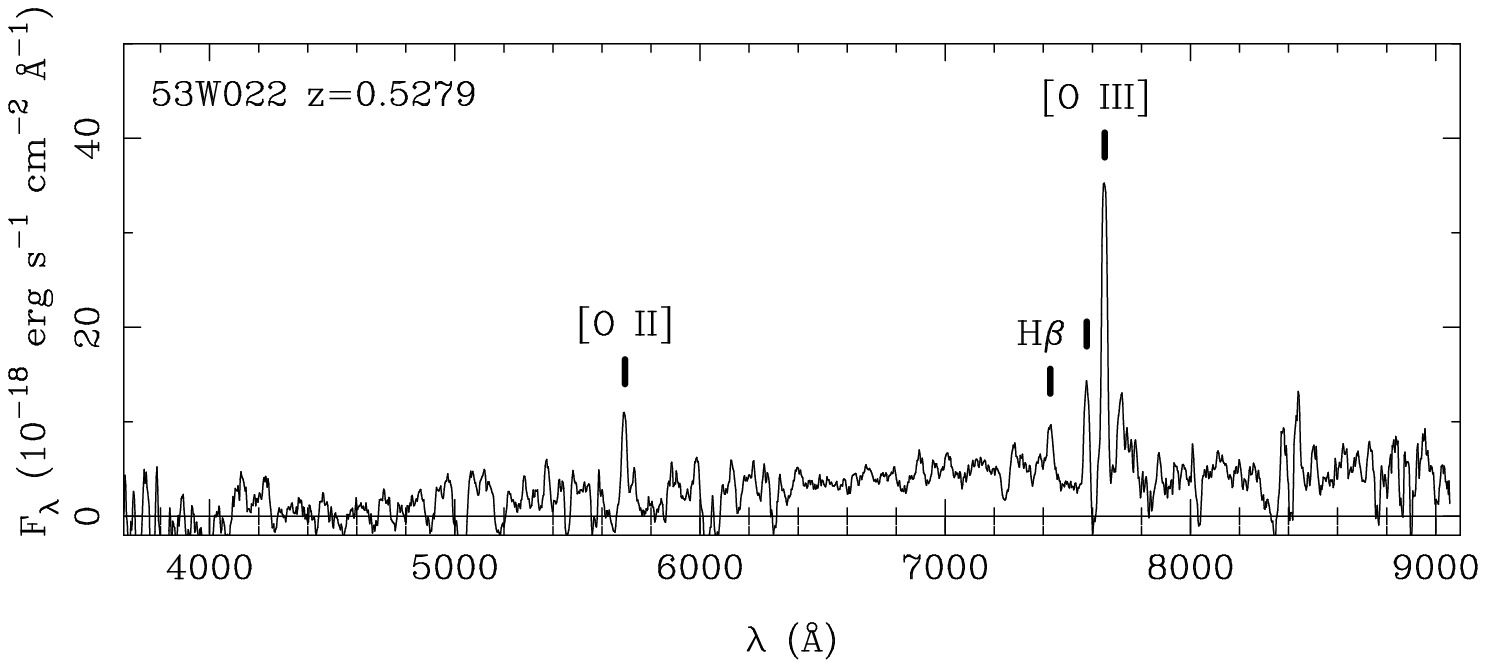,width=8.5cm}\psfig{file=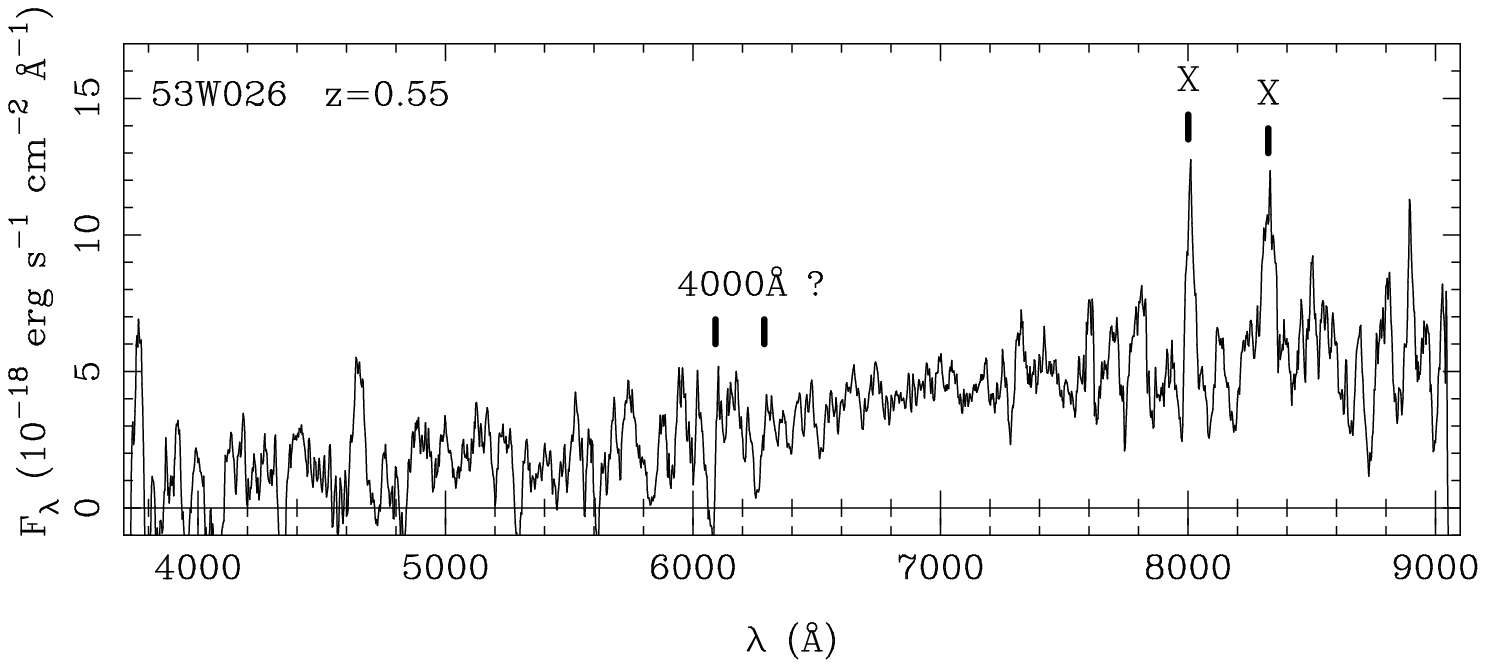,width=8.5cm}}
\hbox{\psfig{file=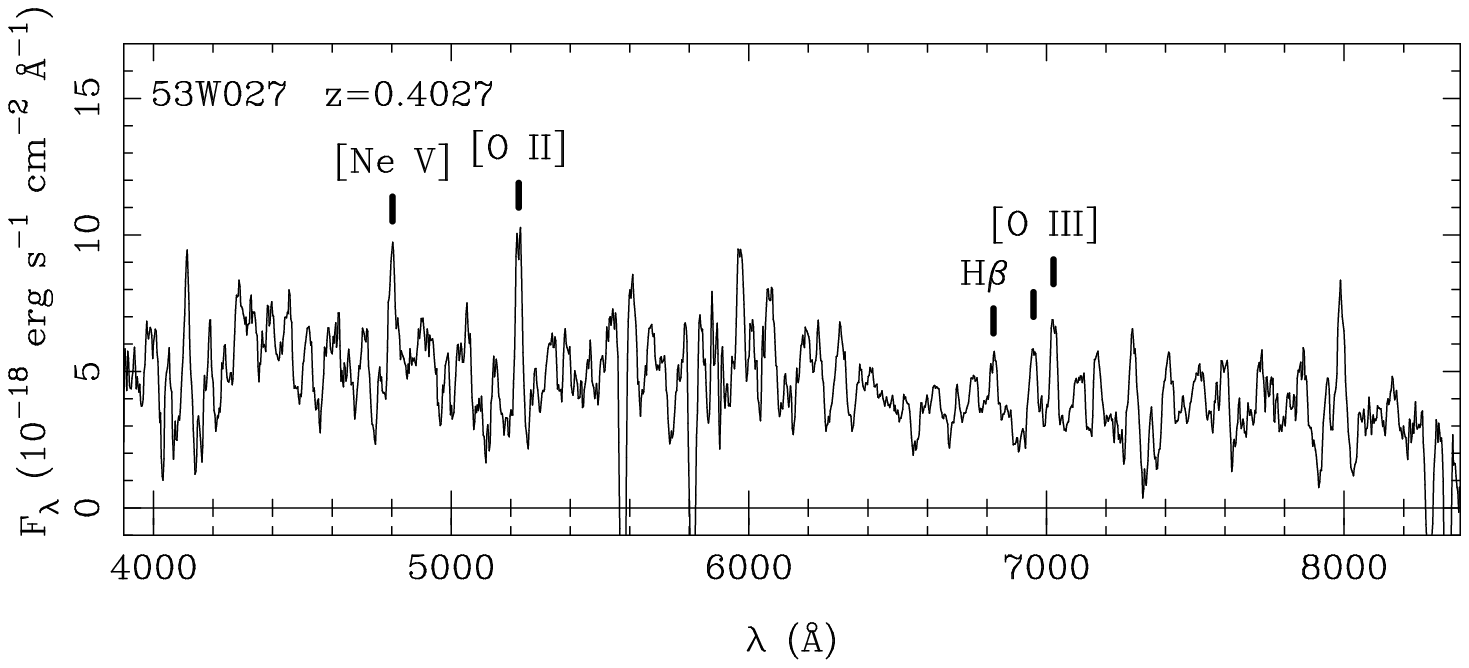,width=8.5cm}\psfig{file=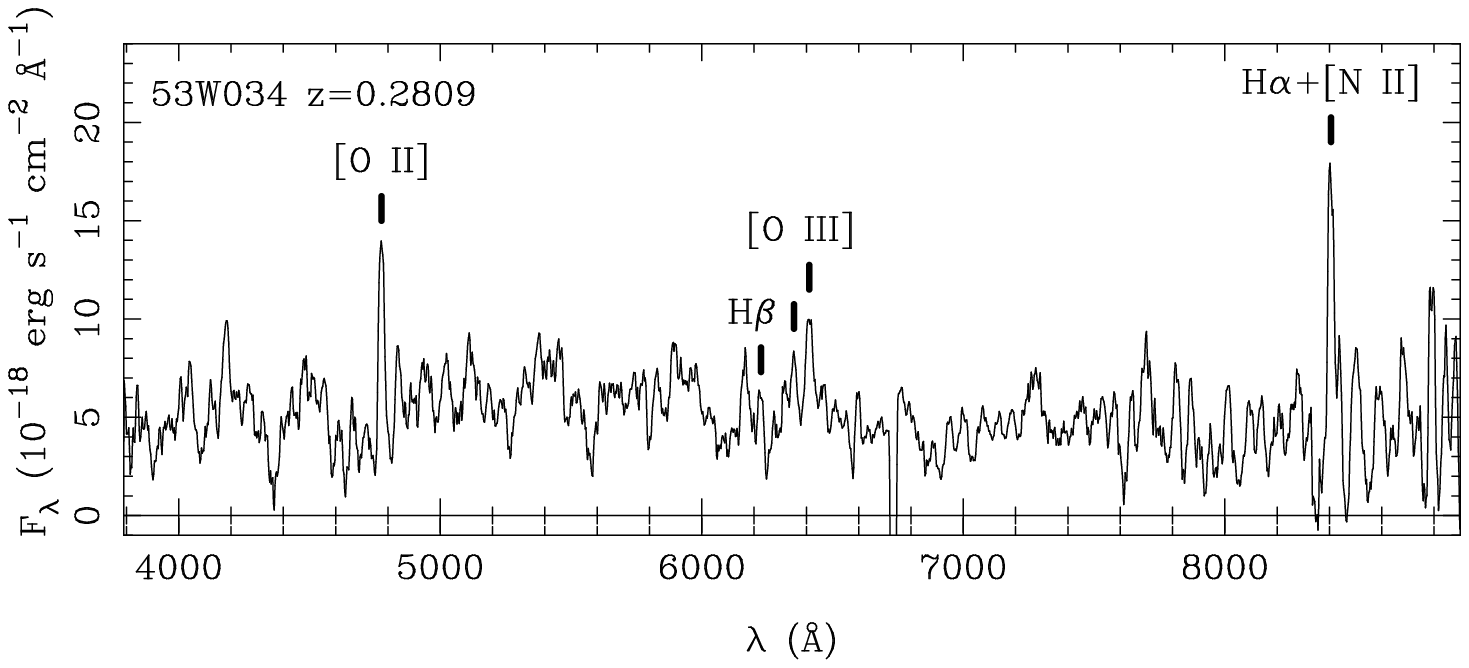,width=8.5cm}}
\hbox{\psfig{file=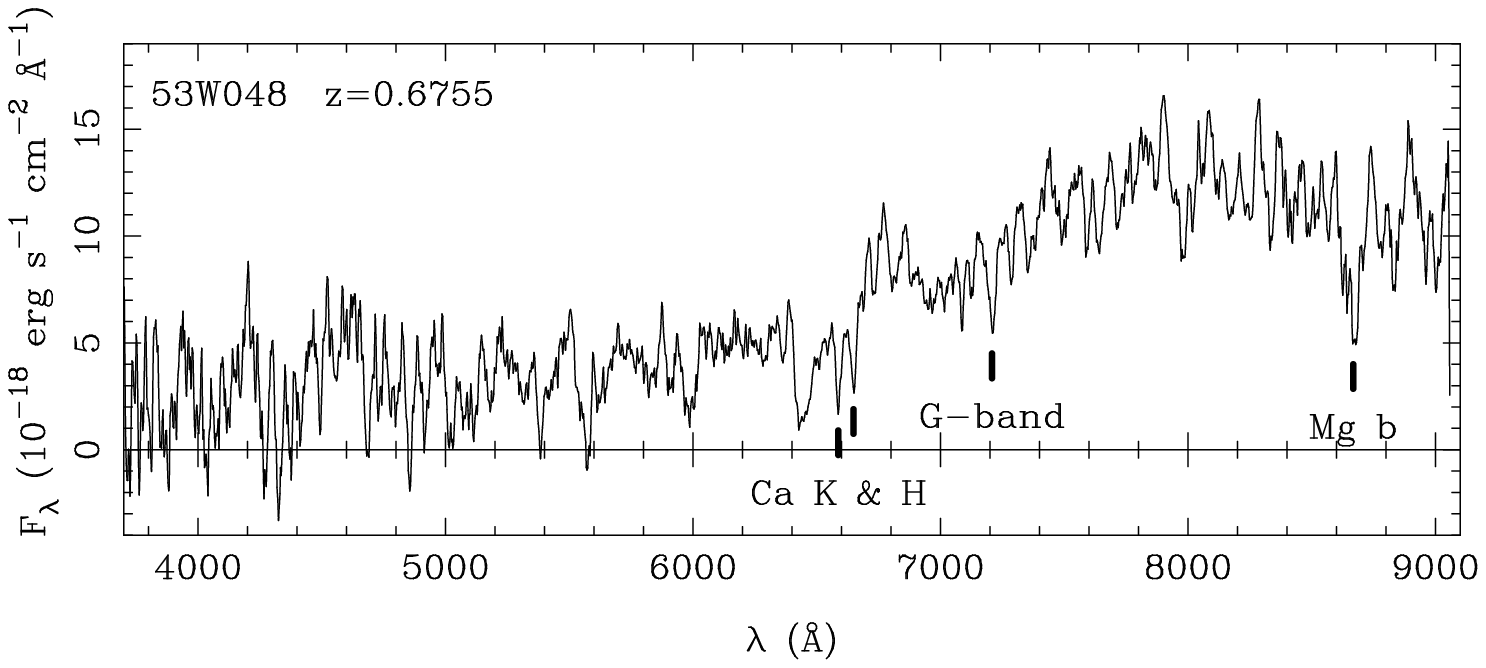,width=8.5cm}\psfig{file=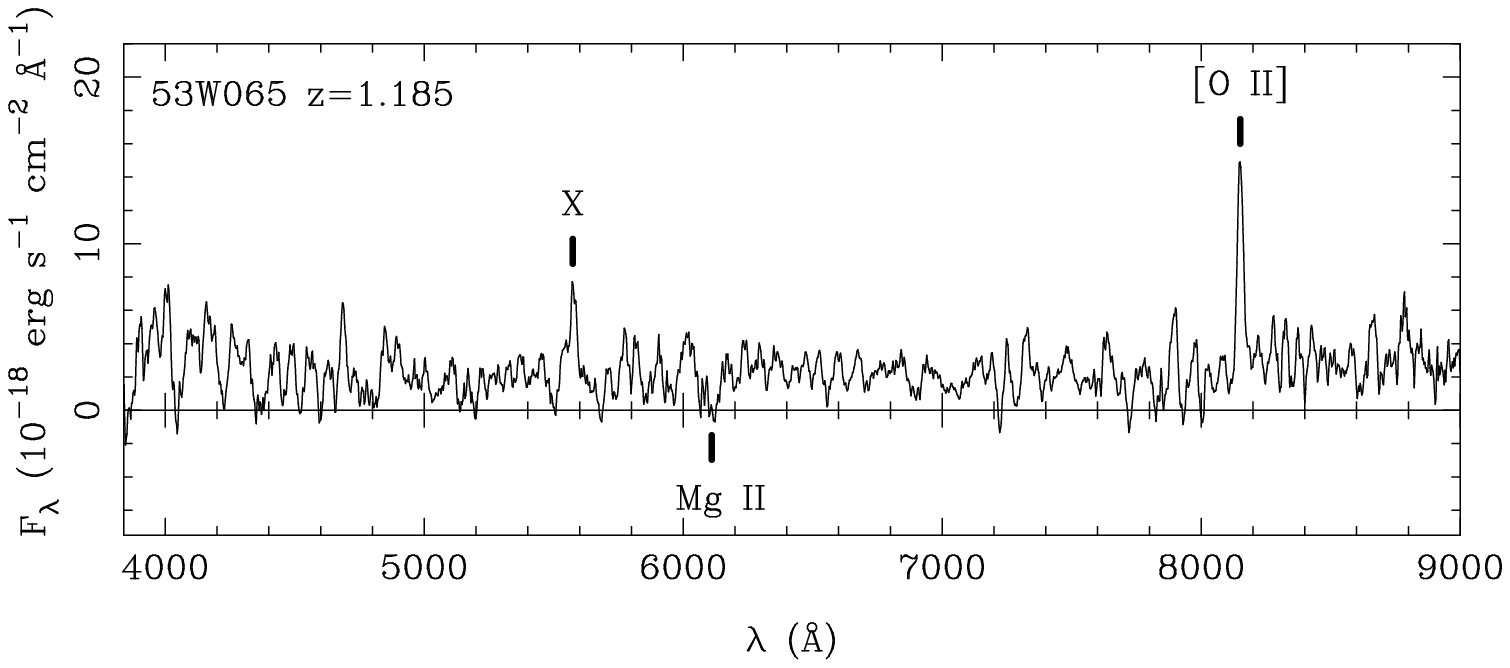,width=8.5cm}}
\hbox{\psfig{file=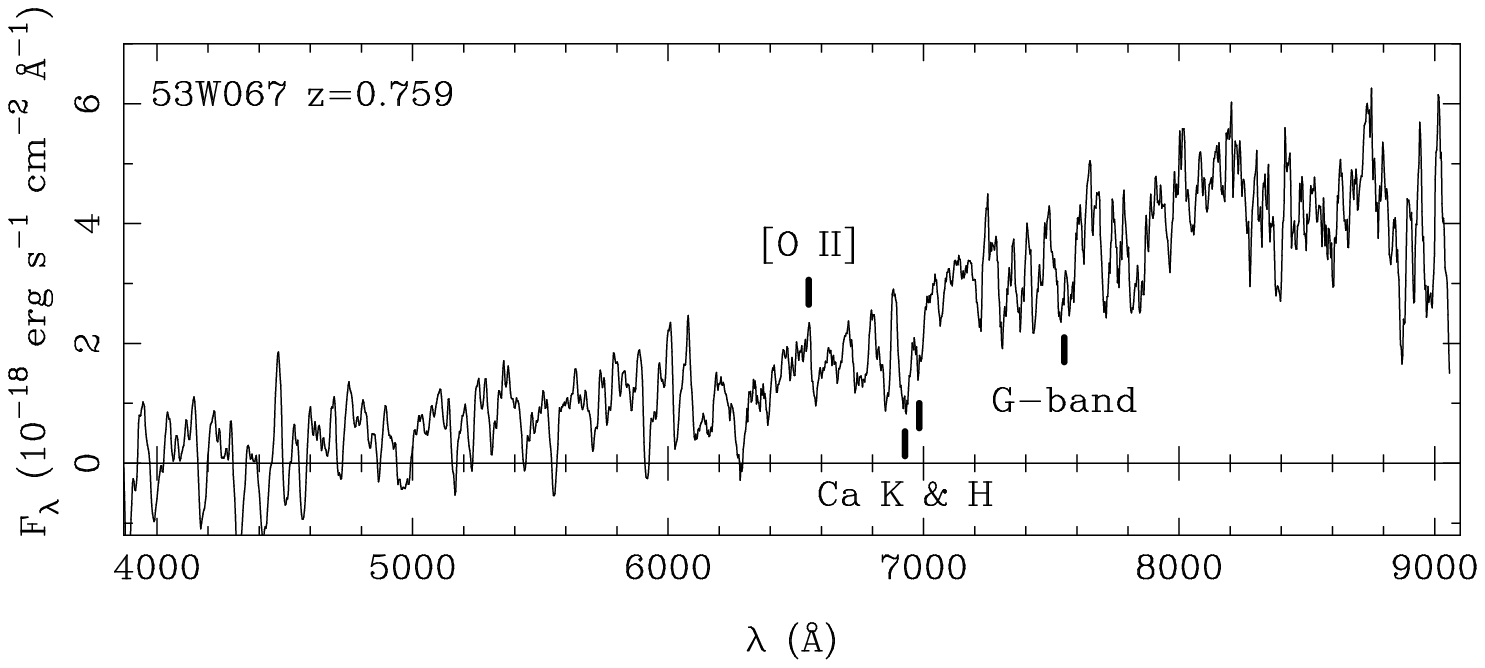,width=8.5cm}\psfig{file=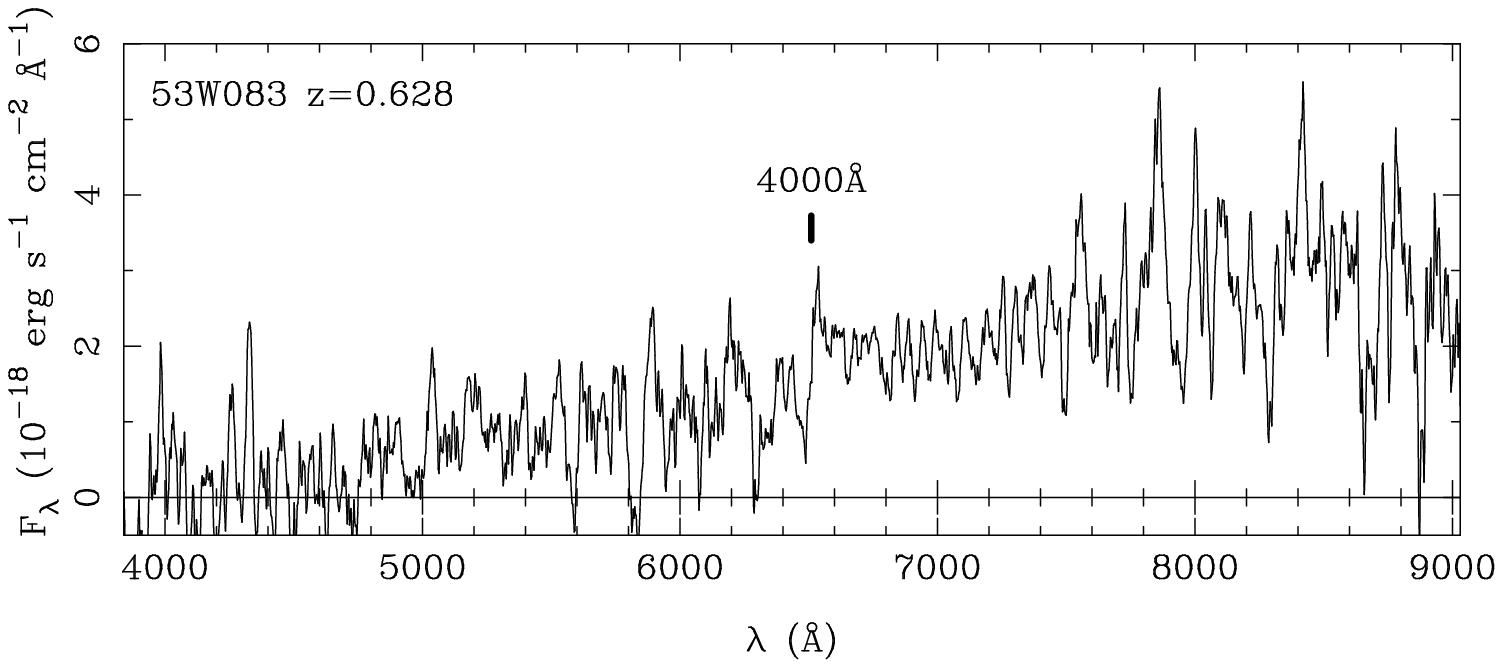,width=8.5cm}}
\caption{WHT spectra of sources in the LBDS Hercules sample for which
a redshift was determined.  The spectra have been smoothed by 32-\AA.
`X' denotes a residual feature such as a night sky line or
incompletely removed CCD fringes.\label{specfig}}
\end{minipage}
\end{figure*}

\begin{figure*}
\begin{minipage}{17cm}
\psfig{file=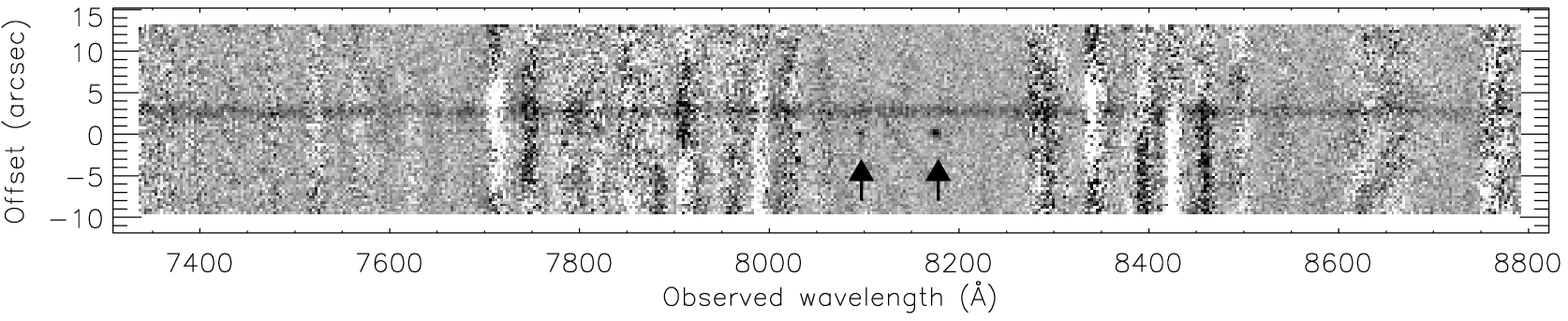,width=17cm}
\caption{Extract from the two-dimensional WHT spectrum of 53W089.  The
abscissa is the dispersion axis, with arrows showing the [O\iii]
4959~\AA\ and 5007~\AA\ emission lines at $z=0.633$.  The ordinate is
distance along the slit (to the west), relative to the position of the
radio source.  The strong continuum emission is from an object
3~arcsec to the west of 53W089.  The spectrum has been sky-subtracted
but not flux-calibrated; residual sky emission lines are seen as broad
vertical bars.\label{specfig89}}
\end{minipage}
\end{figure*}

Of the 28 unique sources observed (one source was observed in both
runs), 11 have yielded a redshift, 10 were detected in continuum
emission but no redshift could be determined, 1 source has a single
emission line but no continuum, and 6 sources were not detected.
Table~\ref{linelist} lists the redshifts and line identifications for
the sources whose redshift has been determined and the spectra are
plotted in figures~\ref{specfig} and \ref{specfig89}.


The redshift of 53W026 has been determined from the shape of the
continuum (c.f.\ 53W019) and two possible identifications of the
4000-\AA\ break.  The location of the redshifted break is uncertain as
it falls within the noisy part of the spectrum where the blue and red
halves were joined, and we assign a redshift of $0.55\pm0.05$
accordingly.  No other features could be consistently identified.  The
[O\ii] emission line in 53W065 is the basis for its redshift of 1.185,
as this is the only strong feature in the spectrum.  There is an
absorption feature at 6110~\AA\ consistent with it being Mg\ii\ at the
same redshift, but this is at the joint between the two halves of the
spectrum and its reality is thus suspect.  53W083 has a very clear
4000-\AA\ break redshifted to 6510~\AA\ putting it at $z=0.628$, but
no other features could be identified in the spectrum.

The only source for which a redshift was determined from its
two-dimensional spectrum was 53W089 at $z=0.634$.  The [O\iii] lines
at restframe wavelengths 4959~\AA\ and 5007~\AA\ are clearly visible
in the sky-subtracted spectrum (figure~\ref{specfig89}).  An [O\ii]
emission line was also detected in the image.  The expected position
of H$\beta$ (usually associated with [O\iii]) is in a region of the
spectrum that is dominated by fringing and the line was not detected.

Ten sources had sufficient continuum flux to enable a strong upper
limit to be placed on the wavelength of the 912-\AA\ Lyman-limit
break.  (Below 912~\AA\ there is essentially no flux in a galaxy's
spectrum due to absorption of the ionizing radiation by H\ione.)  This
enables upper limits to be placed on the redshifts of these sources,
which are listed in table~\ref{speclimits}.  For one of these sources
(53W069) a spectrum was subsequently obtained at the Keck telescope,
yielding a redshift of 1.432 \cite{Dey97,Dunlop99}.  Two of the
objects (53W013 and 53W021) were barely detected, but most had
reasonable flux.  Nevertheless, none of these spectra had any lines or
breaks that could be successfully identified with known spectral
features.


\begin{table}
\caption{Upper limits on the redshifts of sources with continuum flux
but no identified spectral features in the WHT
observations.\label{speclimits}}
\begin{tabular}{ccc} \hline
Source & $\lambda_{\rm min}$ (\AA) & $z$         \\ \hline
53W013 & 3860                      & $< 3.2$     \\
53W014 & 3750                      & $< 3.1$     \\
53W021 & 6000                      & $< 5.6$     \\
53W029 & 4500                      & $< 3.9$     \\
53W035 & 4200                      & $< 3.6$     \\
53W036 & 3800                      & $< 3.2$     \\
53W042 & 3200                      & $< 2.5$     \\
53W068 & 3750                      & $< 3.1$     \\
53W069 & 7200                      & $< 6.9$     \\
53W070 & 3750                      & $< 3.1$     \\ \hline
\end{tabular}
\end{table}

One source (53W060) yielded a possible detection of a single emission
line in the two-dimensional spectrum.  One can speculate that the line
may be either redshifted \Lya\ at $z=2.661$ or redshifted [O\ii] at
$z=0.195$ as these are the strongest lines normally observed in active
galaxies.  With $r\simeq 25.5$~mag the \Lya\ identification is
favoured on the basis of the $r$-band Hubble diagram, however without
independent confirmation this is a very uncertain result and has not
been included in our analysis.

Two of the six spectroscopic non-detections (53W037 and 53W087) do not
have optical counterparts.  We observed them anyway, in the hope that
we could detect an emission line that may have been washed-out in the
broadband observations.  Another of the spectroscopic non-detections
(53W091) was subsequently observed with the Keck telescope and a
redshift of 1.552 was measured \cite{Dunlop96,Spinrad97}.

These results have now brought the total number of redshifts in the
LBDS Hercules sample to 47 out of 72 sources (65\%).  For the 2-mJy
subsample (those sources with $S_{\rm 1.4~GHz}\ge 2$~mJy, see Paper
I), 40 of the 63 sources (63\%) have redshifts.  The redshift content
of the sample is now quite substantial, but our experience has shown
that it will only be completed with the new generation of 8-m class
telescopes.

\section{Photometric redshift estimates}

In order to use the Hercules data to investigate the radio luminosity
function and redshift cut-off, the redshifts of the remaining sources
must be estimated.  The most widely used methods of photometric
redshift estimation involve fitting the data to either observed or
model spectral energy distribution (SED) templates.  A review of the
techniques and the reliability of the results is given by Hogg \etal\
(1998)\nocite{Hogg98}.  In this section we describe the method used to
estimate the redshifts of sources in the LBDS Hercules field, compare
the estimates with the known spectroscopic redshifts and then discuss
the basic properties of the resulting redshift distribution.

\subsection{Estimation method}

In order to develop the best method of calculating photometric
redshifts, the ideas of several authors were combined and
investigated.  Galaxy spectra from the new spectral synthesis models
of Jimenez \etal\ (1998, 2001)\nocite{Jimenez98,Jimenez01} were used
to compute the template SEDs.  We found that the estimated redshifts
were not sensitive to the choice of initial mass function (IMF) or
metallicity, and the spectral synthesis models with Miller-Scalo IMF
and solar metallicity were used in the final analysis.  A template
spectrum was computed by selecting a spectrum of the required age
$F_\lambda^{\rm old}$, then adding to that a blue component
$F_\lambda^{\rm blue}$ scaled such that this component is a fraction
\flilly\ of the total flux at 5000~\AA.  Three forms of the blue
component were tested: a young stellar population of age 0.03~Gyr, one
of age 0.1~Gyr (following Lilly~1989\nocite{Lilly89}), and a power-law
$F_\lambda^{\rm blue} \propto (\lambda/5000)^{-2+\alpha}$ with
$\alpha=0.2$ (following Dunlop \& Peacock~1993\nocite{Dunlop93a}).

Intergalactic absorption due to hydrogen systems along the line of
sight was modelled as a damping of the flux at wavelengths shorter
than \Lya, such that $F_\lambda(\lambda<1216~{\rm \AA}) = F_{\lambda0}
e^{-\tau}$, with $\tau=[(1+z)/5.3]^3$ \cite{Gunn65,Madau96}.  The
Lyman-limit discontinuity, due to the weak stellar emission and strong
interstellar H\ione\ absorption shortward of 912~\AA, was modelled by
a cut-off of the form $F_\lambda(\lambda<912~{\rm \AA})=0$.  The
effects of possible dust absorption either within the galaxy or along
the line of sight were not considered -- the additional parameters
could not be justified, given that each source had on average only
four flux measurements for comparison with the models.

The synthetic flux in each filter ($U^+griJHK$) was computed by
convolving the model spectrum with the appropriate filter response
function (computed from the filter transmission, the atmospheric
extinction and the quantum efficiency of the CCD detector).  The
fluxes were then converted to magnitudes using tabulated flux density
measurements of the standard stars BD$+$17$\degr$4708 ($gri$ filters)
and Vega ($U^+JHK$ filters).  Synthetic magnitudes were generated for
redshifts $0 \le z \le 5$ in steps of 0.1, blue component fractions $0
< f_{5000} < 1$ in steps of 0.02, and twenty galaxy ages
(0.01--14~Gyr).  The set of synthetic magnitudes was compared with the
observed magnitudes of each source, using the \chisq\ statistic.
Non-detections were incorporated into \chisq\ by using the upper
limits to their fluxes.  For every source, \chisq\ was evaluated for
each combination of the parameters $z$, \flilly, and the age, yielding
values of \chisq($z$,\flilly,age) for each of the three different blue
components.  The best-fitting parameters and their errors were
determined by using the Levenberg-Marquardt minimization technique
described in \S15.5 of Press \etal\ (1992)\nocite{Press92}.  To the
extent that the measurement errors are normally distributed, the
errors in the fit are then the 68\% confidence limits on each
parameter separately.

\subsection{Comparison with spectroscopic redshifts}

Since two-thirds of the Hercules sample have spectroscopic redshifts
($z_{\rm spec}$), the best-fit photometric redshift ($z_{\rm phot}$)
can be compared with the true redshift for these sources.  This was
done by directly comparing $z_{\rm phot}$ with $z_{\rm spec}$ for each
source, and secondly, by computing the median, $\langle \Delta z
\rangle_{\rm med}$, of $\Delta z = z_{\rm phot} - z_{\rm spec}$
together with the standard deviation about the median, $\sigma_{\rm
med}$.  Any sources with $\Delta z > 2.5\, \sigma_{\rm med}$ were
rejected and $\sigma_{\rm med}$ was recomputed.  Each redshift
estimation method was assessed on the basis of the values of $\langle
\Delta z \rangle_{\rm med}$ and $\sigma_{\rm med}$, the number of
points rejected ($N_{rej}$) in the second iteration of $\sigma_{\rm
med}$, and the proportion of redshifts that were underestimated (since
it is the underestimated redshifts that would affect the subsequent
analysis of the redshift cut-off most severely).

It became clear from these comparisons that minimizing \chisq\ over
the whole of the parameter ranges produced a very poor result, so we
determined which {\it restricted\/} set of parameters produced the
best redshift estimates.  Several methods were tested. (i) In order to
approximate the methods of the HDF groups (e.g.\ Lanzetta et~al.~1996;
Sawicki et~al.~1997)\nocite{Lanzetta96,Sawicki97}, the number of
spectral templates was limited to three, corresponding to E/S0, Sabc
\& Sd/Irr galaxies.  (ii) Following Lilly (1989)\nocite{Lilly89}, the
old component was fixed at ages of either 5, 10, 12 or 14~Gyr.  (iii)
The age parameter at each redshift was limited such that it was less
than the age of the universe.  (iv) The blue component was restricted
to values of $f_{5000}<0.67,\ <0.5$ and $f_{5000}=0$, to exclude the
most blue SEDs.


\begin{figure}
\psfig{file=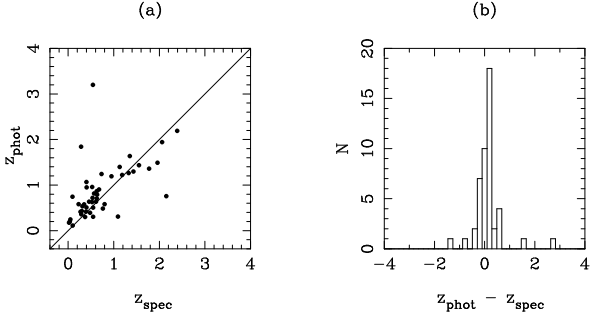}
\caption{Comparison of the photometric estimates $z_{\rm phot}$ with
the spectroscopic redshifts $z_{\rm spec}$ for the 47 sources with
known redshifts.  (a) Direct comparison between estimated and
spectroscopic redshifts.  (b) Histogram of their differences.  The
median difference is $\langle \Delta z \rangle_{\rm med} = 0.13$ and
the standard deviation about this median is $\sigma_{\rm med} =
0.29$.\label{herczest}}
\end{figure}

Of these methods, it was the second one that produced the most
accurate results (figure~\ref{herczest}) -- fixing the age of the old
component and fitting to $z$ and \flilly.  Using the maximum available
age for the old stellar population (14~Gyr) and taking the blue
component to be a 0.1~Gyr stellar population, gave $\langle \Delta z
\rangle_{\rm med} = 0.13$ and $\sigma_{\rm med} = 0.29$.  Note that
the median difference between photometric and spectroscopic redshifts
remains well within the error distribution.


\begin{figure}
\psfig{file=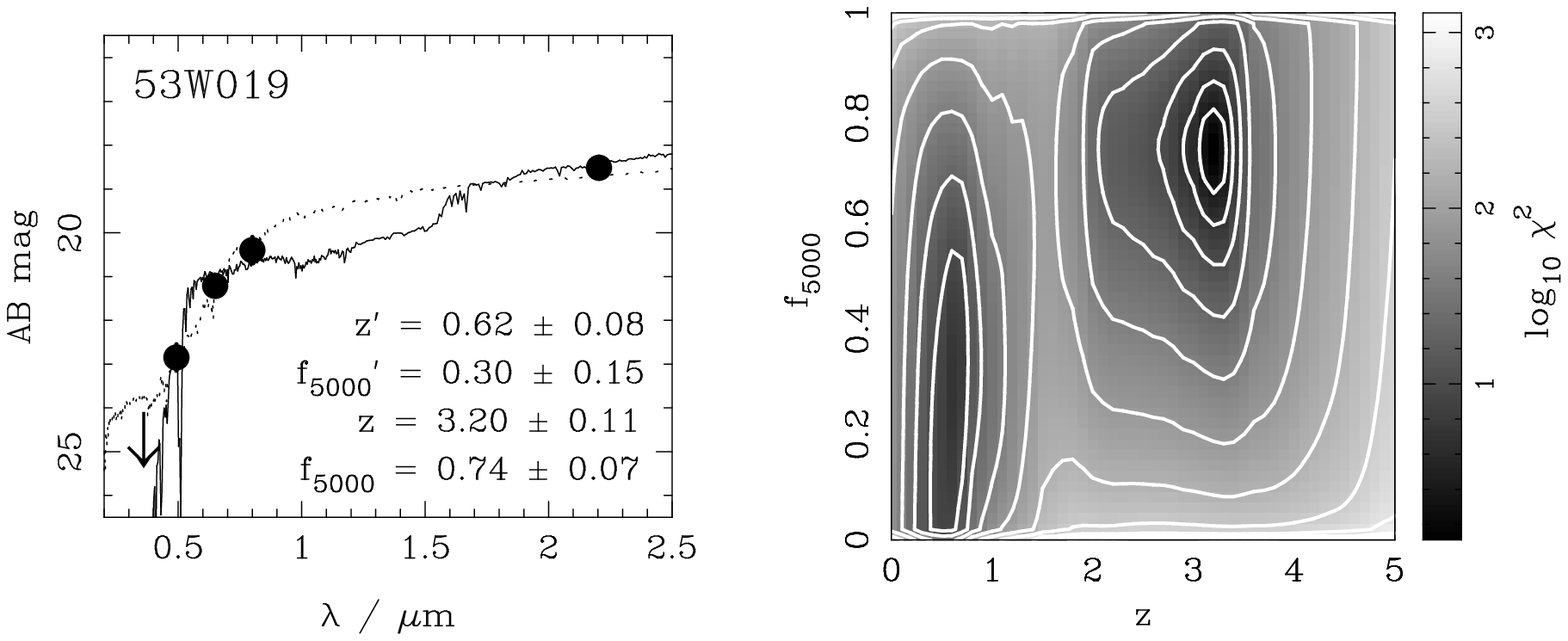,width=8.5cm}
\psfig{file=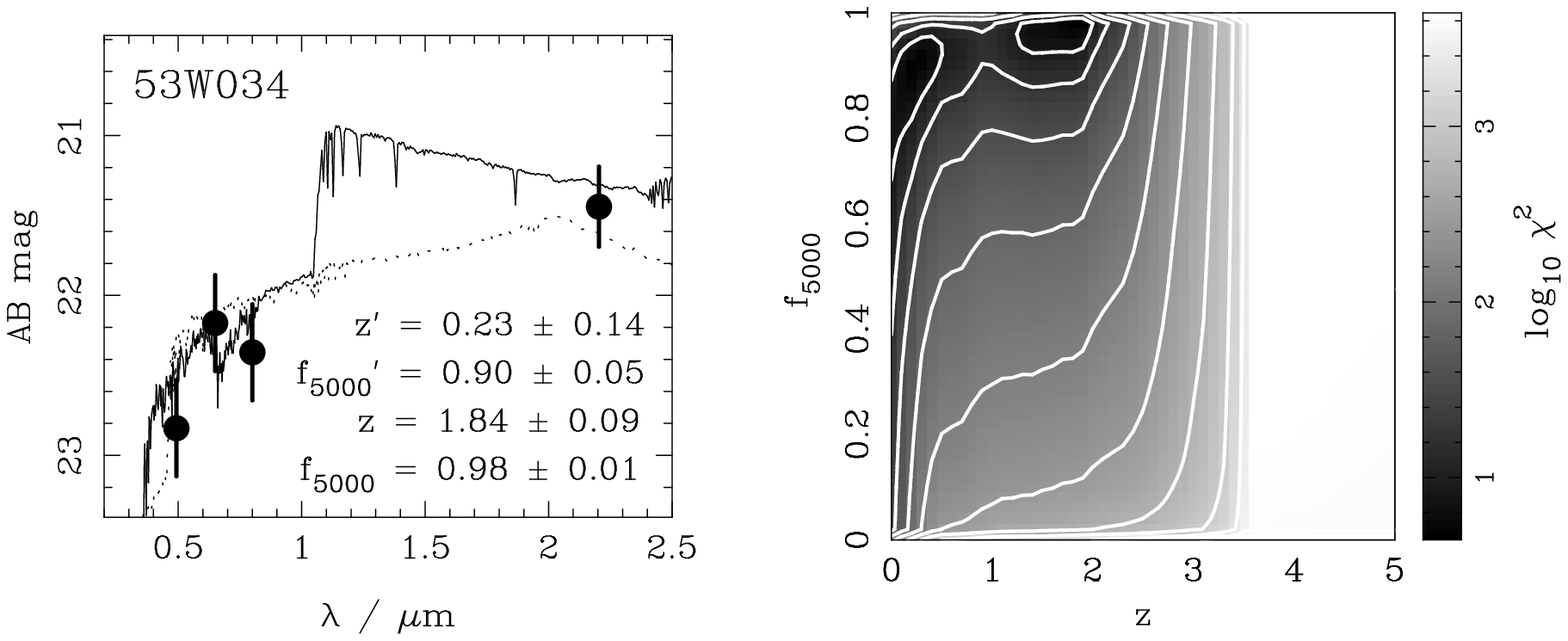,width=8.5cm}
\caption{Redshift estimates for 53W019 ({\it top\/}) and 53W034 ({\it
bottom\/}).  Solid lines denote the best-fitting photometric redshift
($z$), dotted lines denote the redshift ($z^{\prime}$) corresponding
to the secondary minimum in \chisq.  The spectroscopic redshifts are
0.542 and 0.281 respectively.\label{badzphots}}
\end{figure}

It can be seen in figure~\ref{herczest} that there are several sources
with small spectroscopic redshifts ($z_{\rm spec}\la 0.5$) that are
assigned significantly larger photometric redshifts.  This effect can
also be seen in the work of Lanzetta \etal\ (1996)\nocite{Lanzetta96}
and Ellis (1997)\nocite{Ellis97}.  In the case of 53W090 at $z_{\rm
spec}=0.094$, this is probably due to poor photometry -- it is a
bright galaxy that extends beyond the apertures used in both the
optical and IR images.  For the other sources, there appear to be two
effects: (i) for $z_{\rm phot} \sim 2$ the location of the \fourang\
can be anywhere from 1~\micron\ to 2~\micron\ due the lack of infrared
data at $J$ \& $H$; and (ii) for $z_{\rm phot} \sim 3$ the \fourang\
is being mis-interpreted as the Lyman-limit break at 912~\AA.  In
virtually all cases a second minimum in \chisq\ is found close to the
spectroscopic redshift.  Figure~\ref{badzphots} illustrates this for
53W019 and 53W034, the two spectroscopic sources which clearly show
this effect in figure~\ref{herczest}(a).  Both these sources have a
second minimum in \chisq\ that differs from their true redshift by
$\Delta z<0.1$.

The two sources that have photometric redshifts that are significantly
underestimated by this method are 53W009 ($z_{\rm spec} = 1.090$) and
53W075 ($z_{\rm spec}= 2.150$).  Both are classified as quasars and it
is perhaps not too surprising that their SEDs are not successfully
modelled by a galaxy spectrum.  Note, however, that the redshifts of
the other four quasars in the sample are correctly estimated.

\subsection{Results}

The best-fitting model SEDs and the \chisq\ functions for the
twenty-two sources without spectroscopic redshifts are shown in
figure~\ref{photzplots}.  Figure~\ref{herclikelihood} presents the
likelihood functions $L(z)$ for these sources, normalized to a peak
value of unity.  The likelihood is defined as $L(z) \propto {\rm exp}(
- \chi^2(z) / 2 )$, where \chisq\ has been minimized with respect to
\flilly, i.e.\ $\chi^2(z)=\chi^2(z,f_{\rm min})$ where $f_{\rm min}$
is the value of \flilly\ for which \chisq\ is a minimum at each fixed
$z$.  Table~\ref{photztable} lists the best-fitting photometric
redshift for each source, together with the one-sigma error
$\sigma_z$, the blue component fraction \flilly\ and its error
$\sigma_f$, and the \chisq\ of the fit.  There were typically 2--4
degrees of freedom for each fit, thus it can been seen that most of
the \chisq\ values indicate reasonable fits to the data.  Two of the
sources (53W051 \& 53W070) have a minimum \chisq\ significantly
greater than that required for a good fit, but in both cases this can
be attributed to the small photometric errors in the data, rather than
any uncertainty in the estimated redshift.


\begin{figure*}
\begin{minipage}{175mm}
\hbox{%
\psfig{file=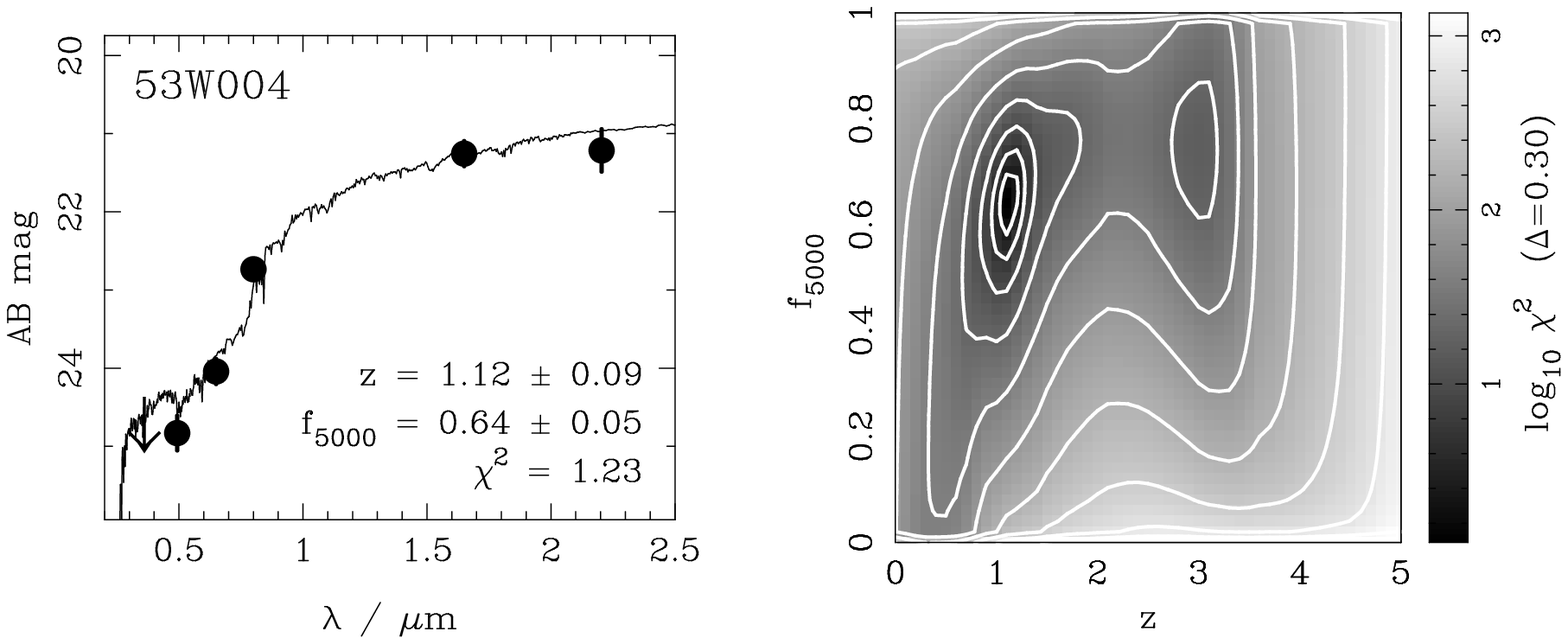,width=8cm}\ \ \ \ \ \ 
\psfig{file=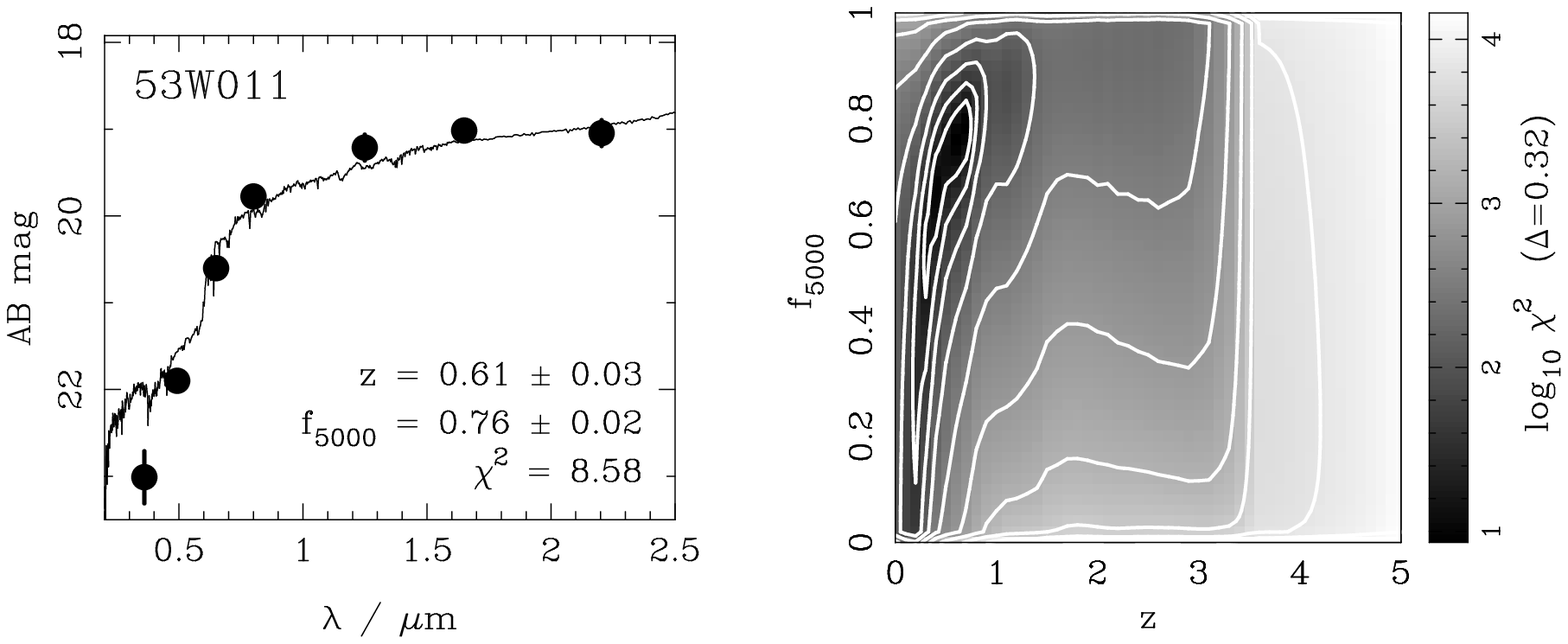,width=8cm}}
\hbox{}
\hbox{%
\psfig{file=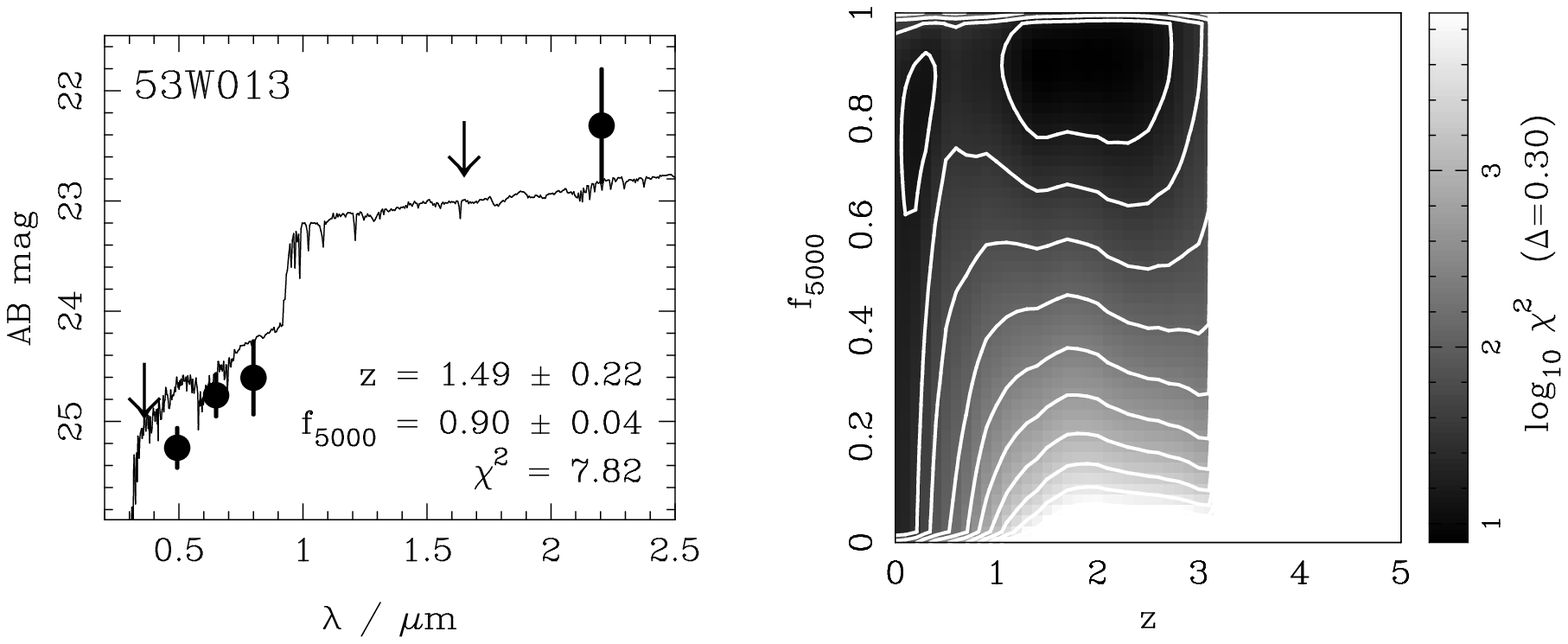,width=8cm}\ \ \ \ \ \ 
\psfig{file=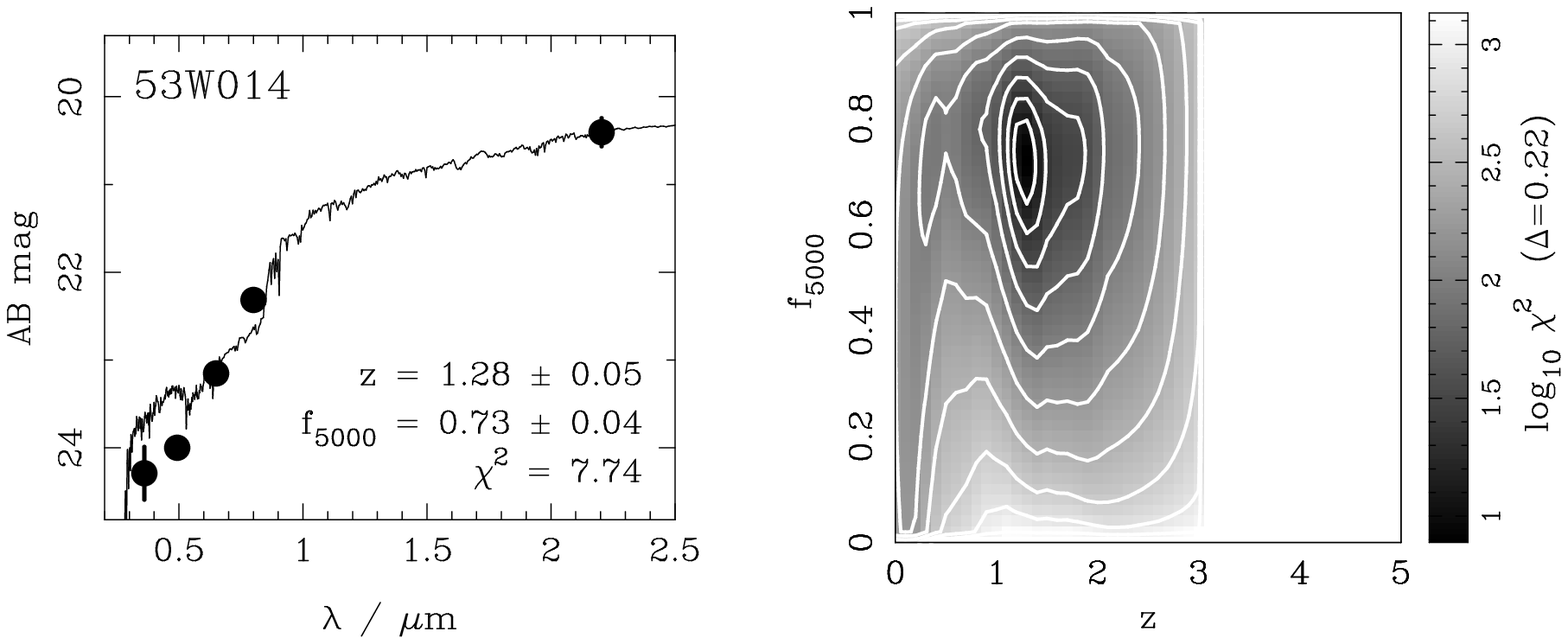,width=8cm}}
\hbox{}
\hbox{%
\psfig{file=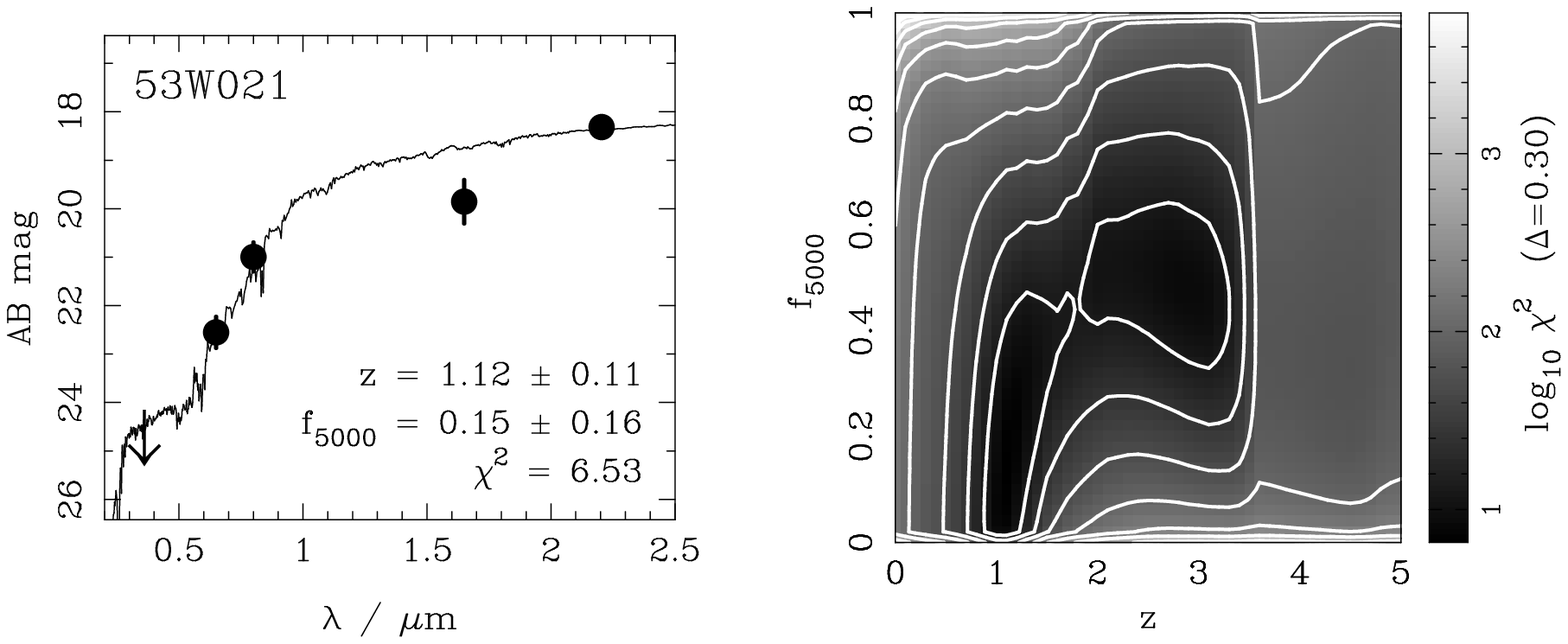,width=8cm}\ \ \ \ \ \ 
\psfig{file=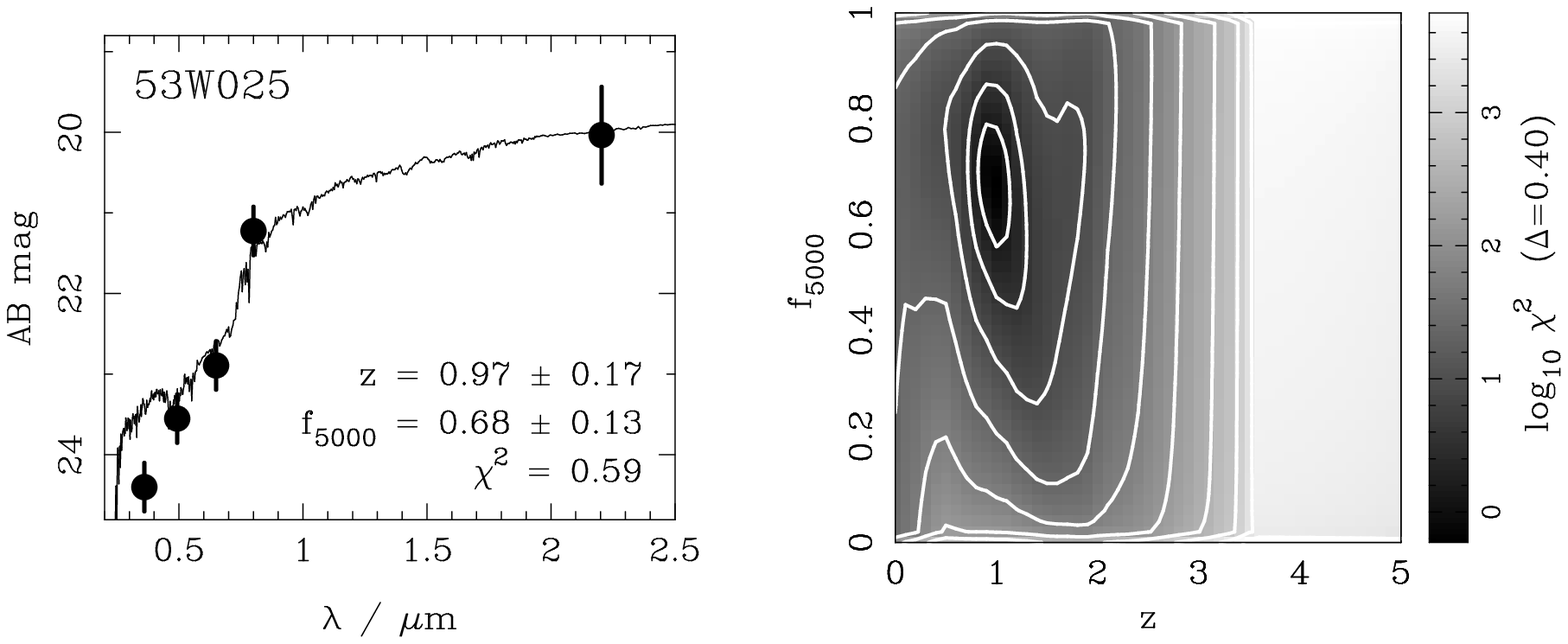,width=8cm}}
\hbox{}
\hbox{%
\psfig{file=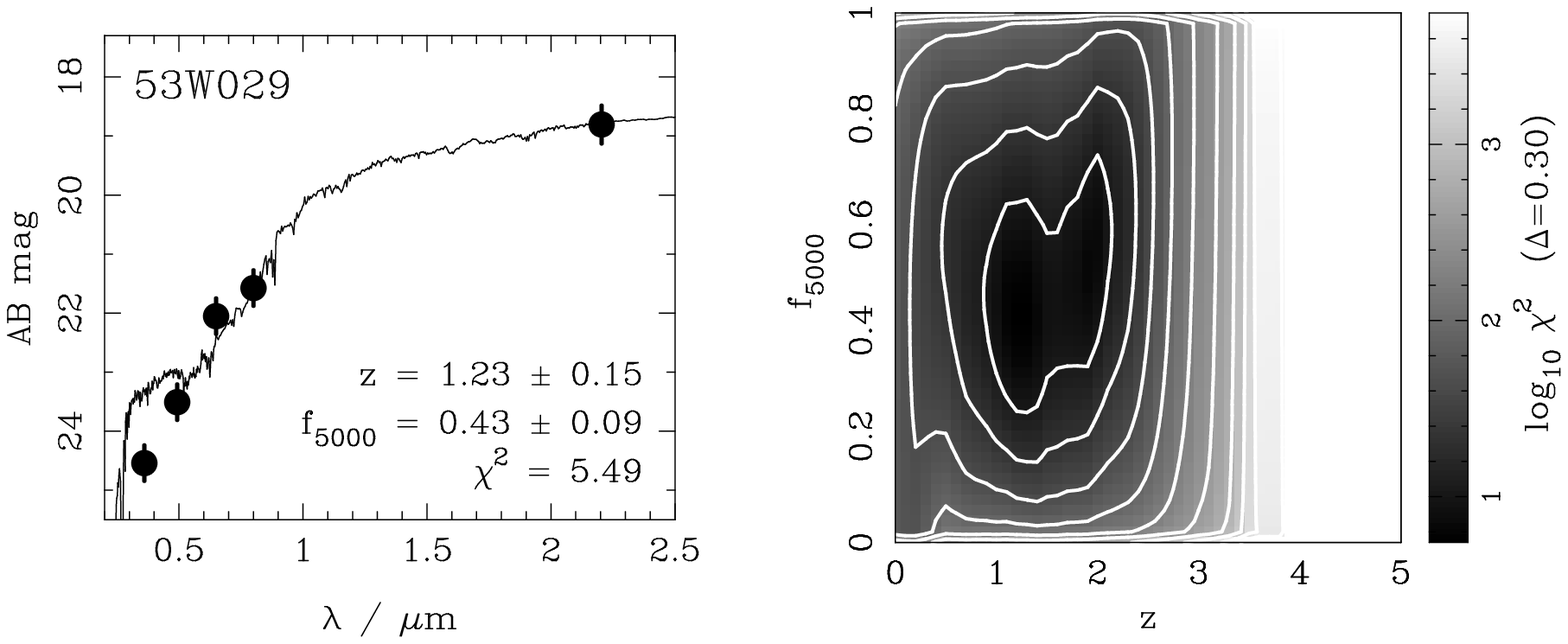,width=8cm}\ \ \ \ \ \ 
\psfig{file=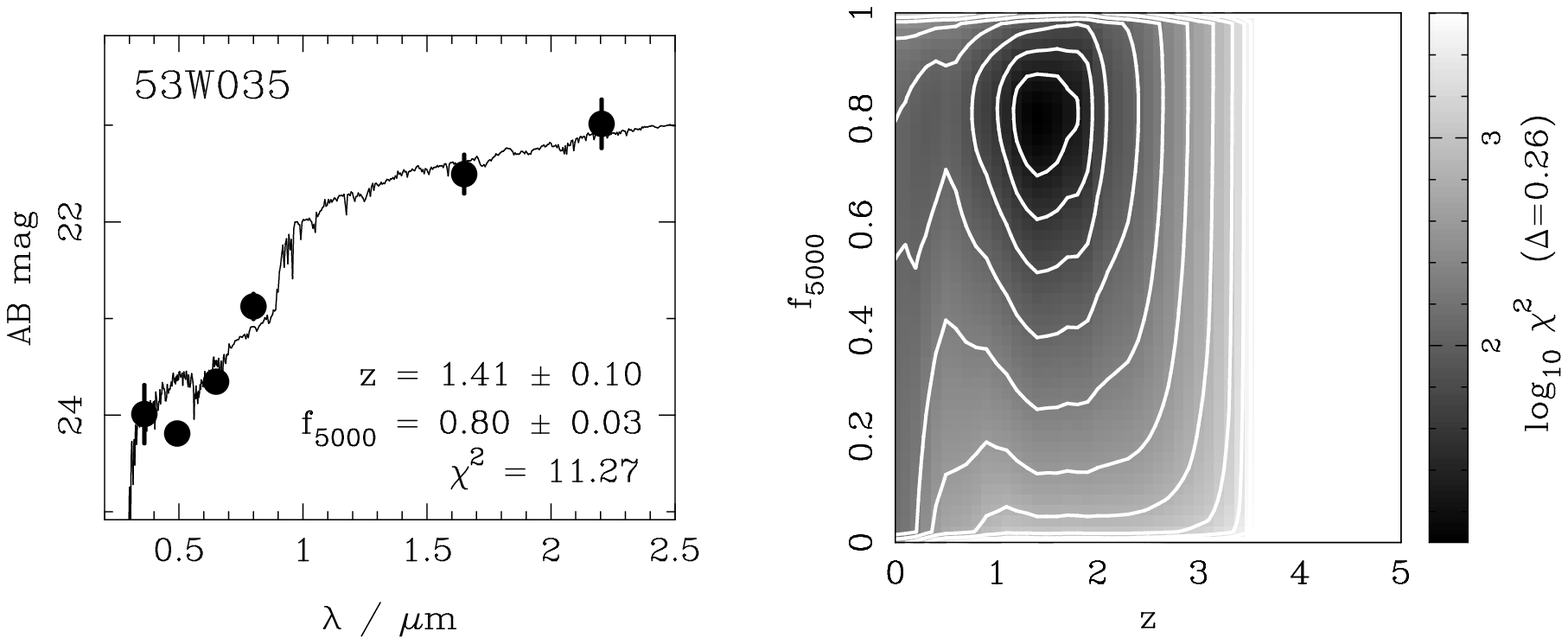,width=8cm}}
\hbox{}
\hbox{%
\psfig{file=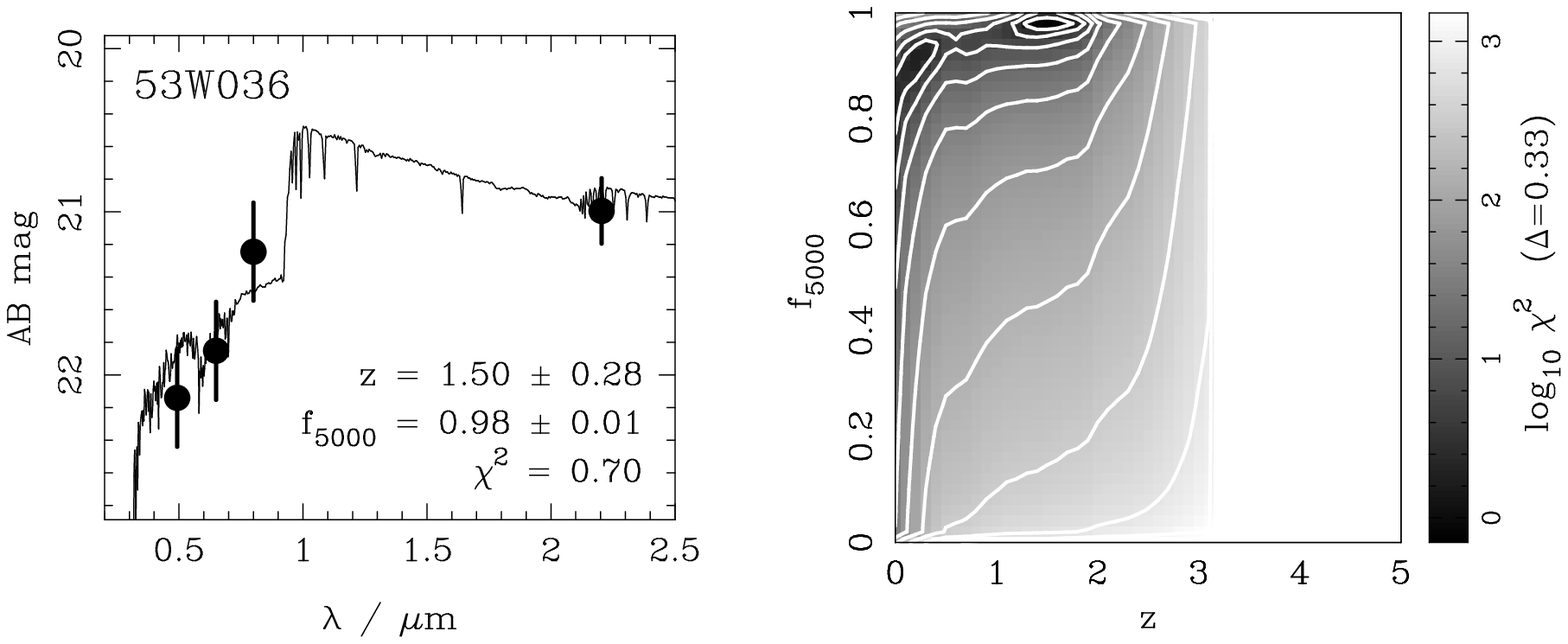,width=8cm}\ \ \ \ \ \ 
\psfig{file=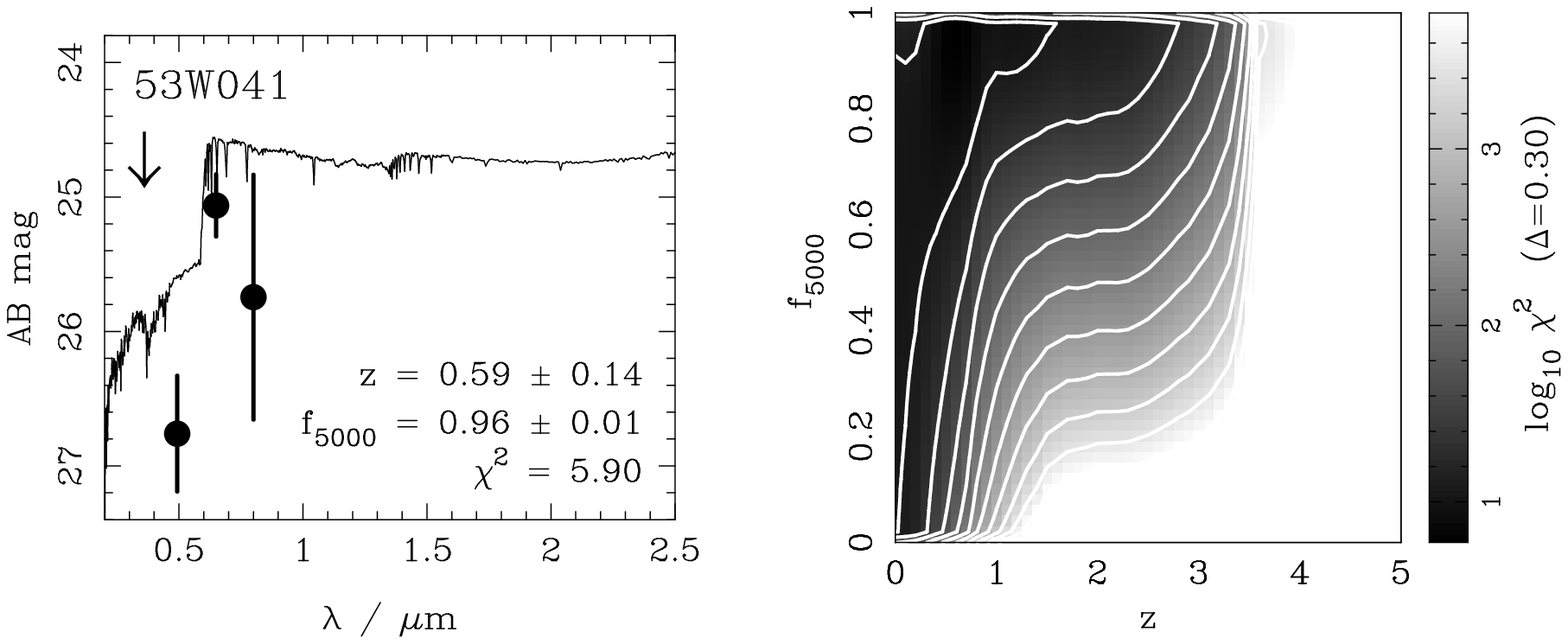,width=8cm}}
\hbox{}
\hbox{%
\psfig{file=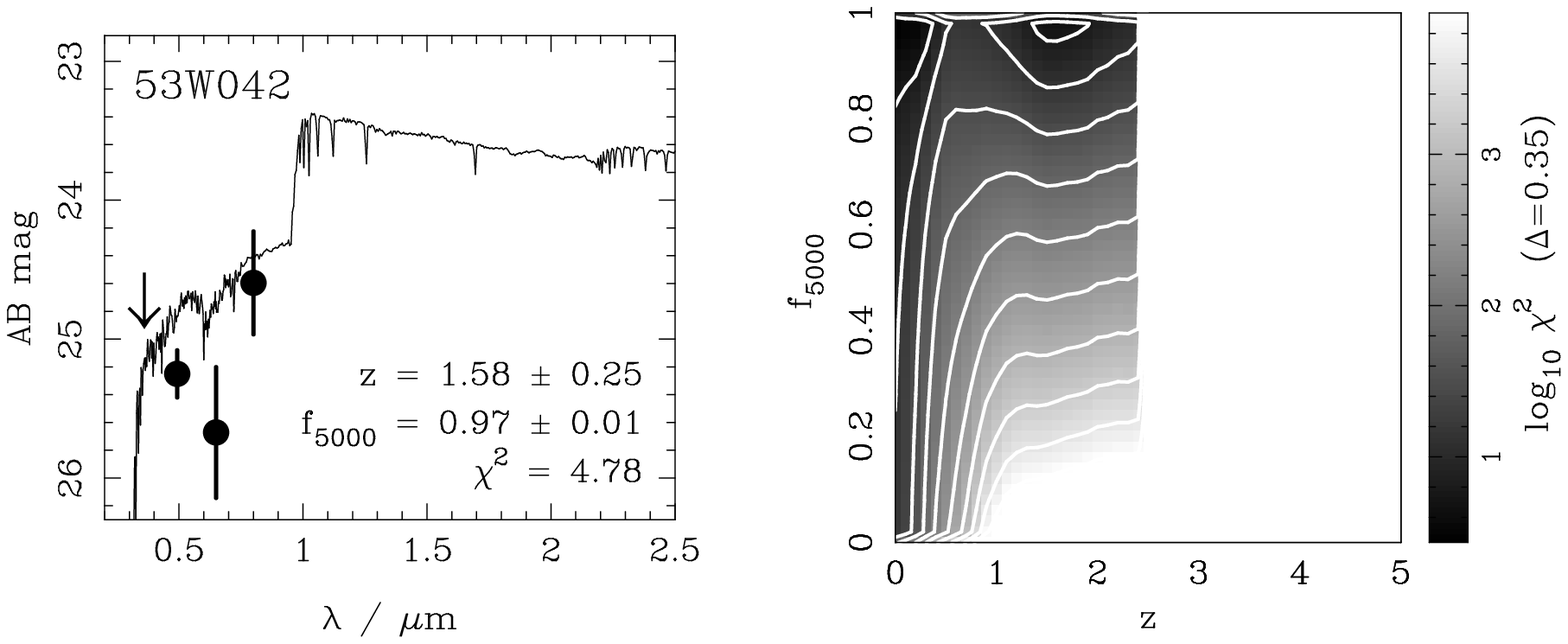,width=8cm}\ \ \ \ \ \ 
\psfig{file=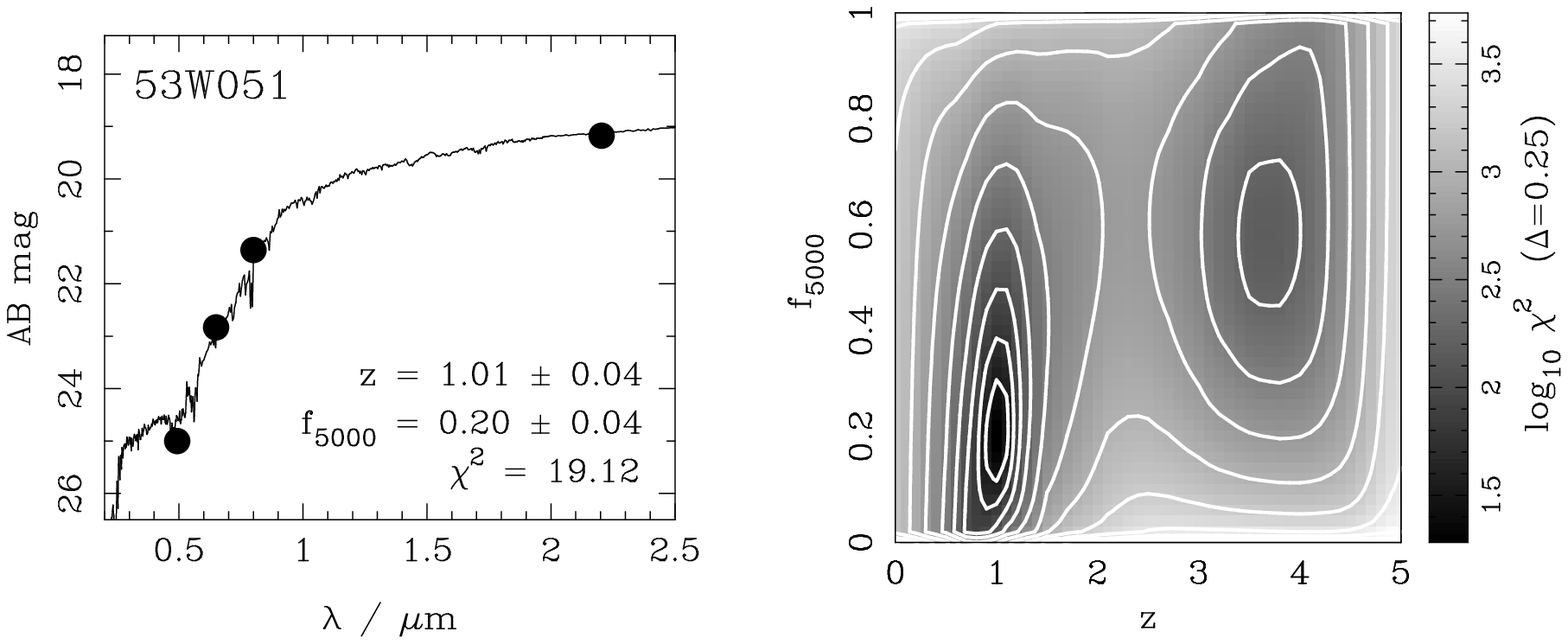,width=8cm}}
\caption{The twenty-two sources in the Hercules sample for which
photometric redshifts have been determined.  For each source, the
best-fitting model SED is shown on the left, together with the
observed data points (converted to AB magnitudes); on the right, we
plot $\logten \chi^2$ as a function of $z$ and \flilly.  The step size
between successive logarithmic contours is given in parentheses
($\Delta$).\label{photzplots}}
\end{minipage}
\end{figure*}

\begin{figure*}
\begin{minipage}{175mm}
\hbox{%
\psfig{file=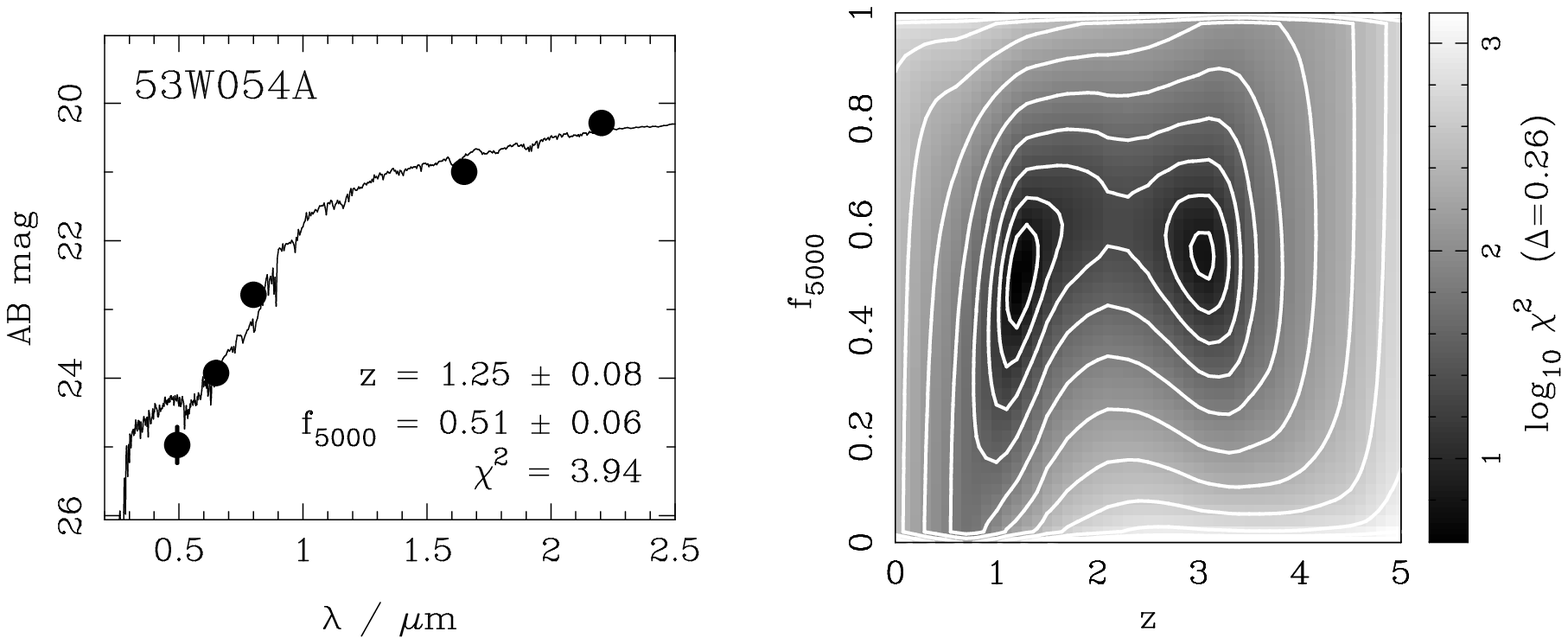,width=8cm}\ \ \ \ \ \ 
\psfig{file=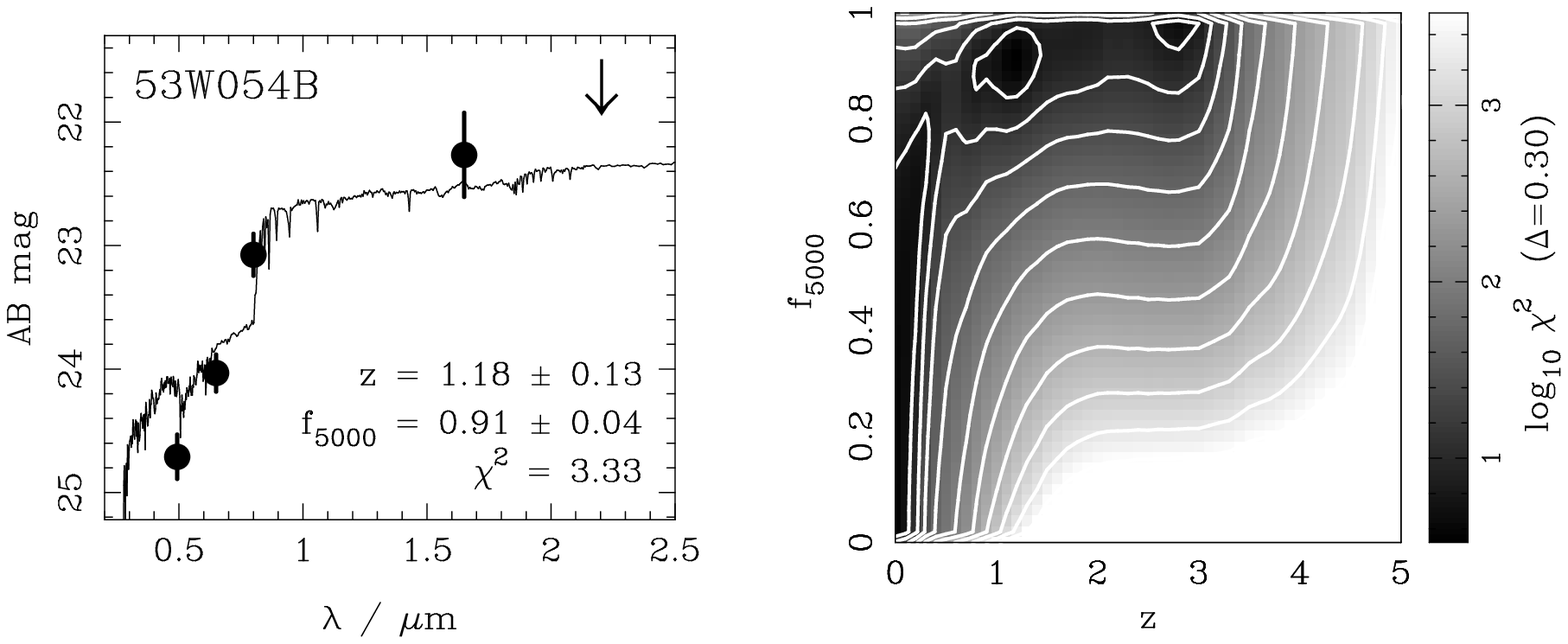,width=8cm}}
\hbox{}
\hbox{%
\psfig{file=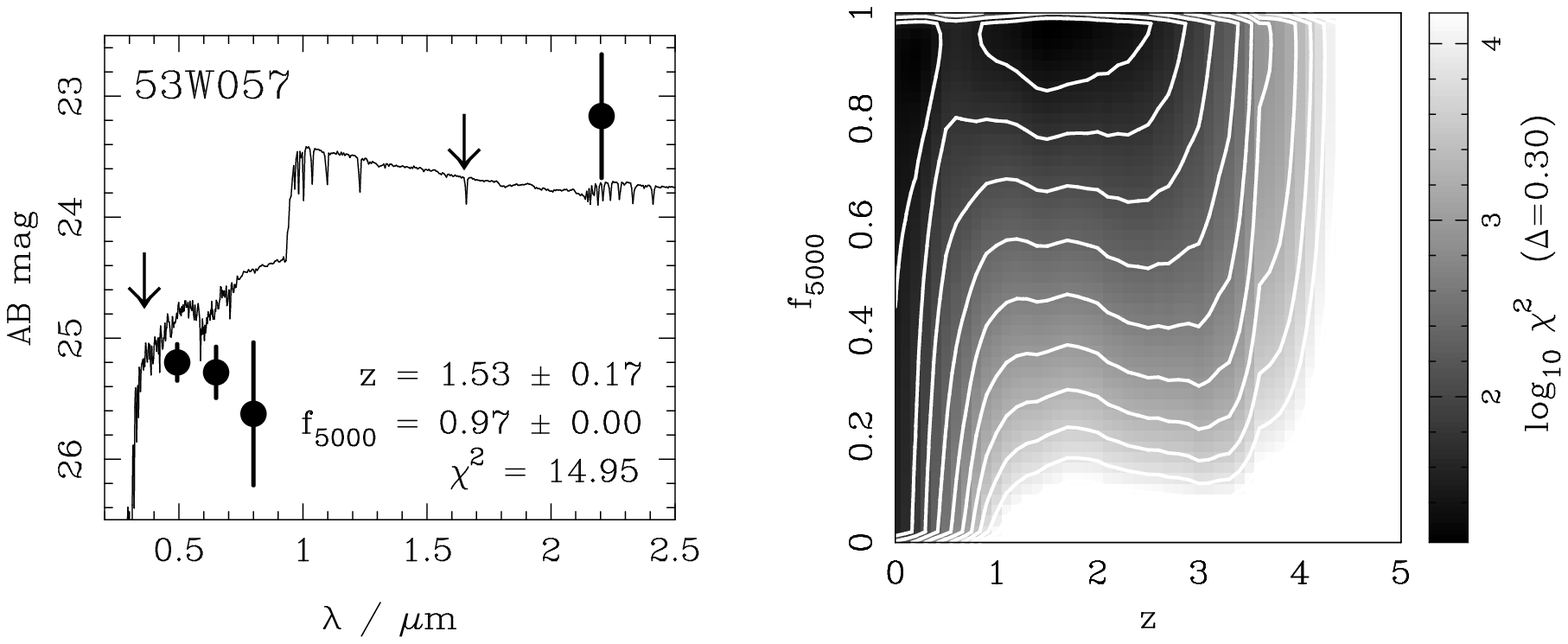,width=8cm}\ \ \ \ \ \ 
\psfig{file=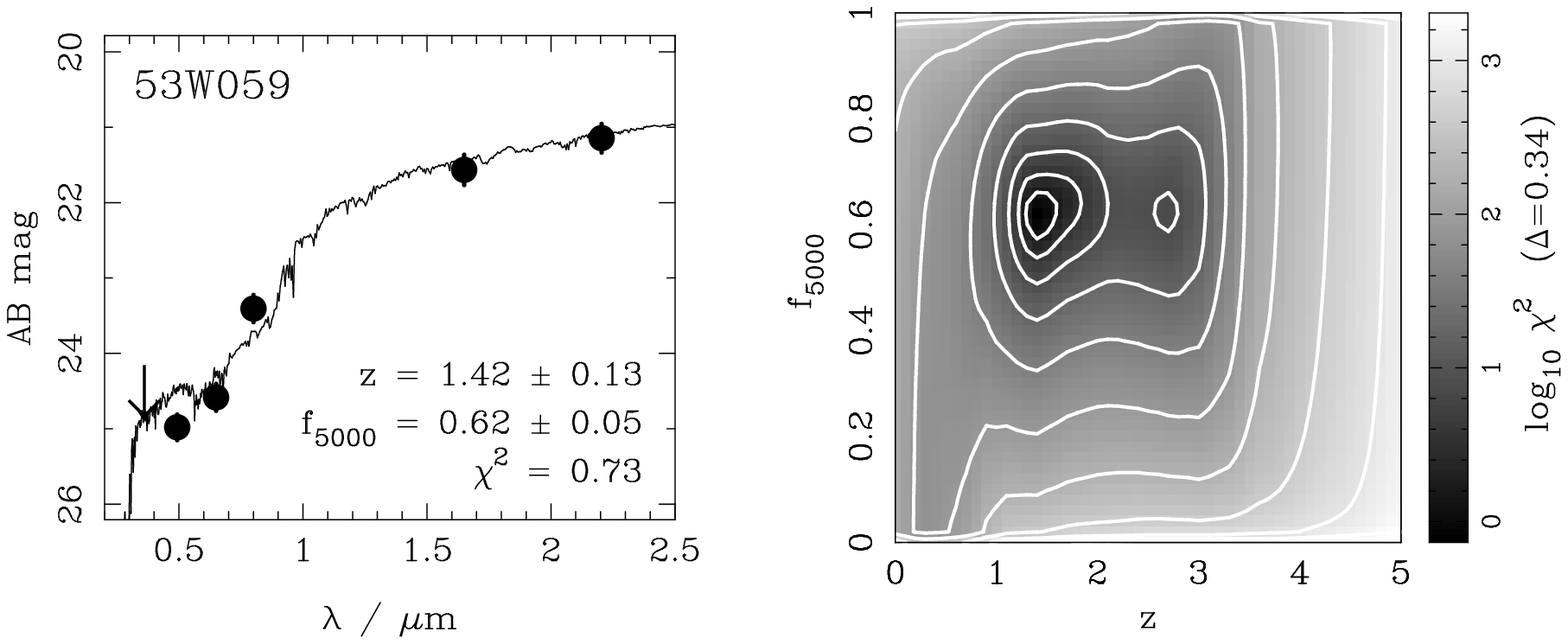,width=8cm}}
\hbox{}
\hbox{%
\psfig{file=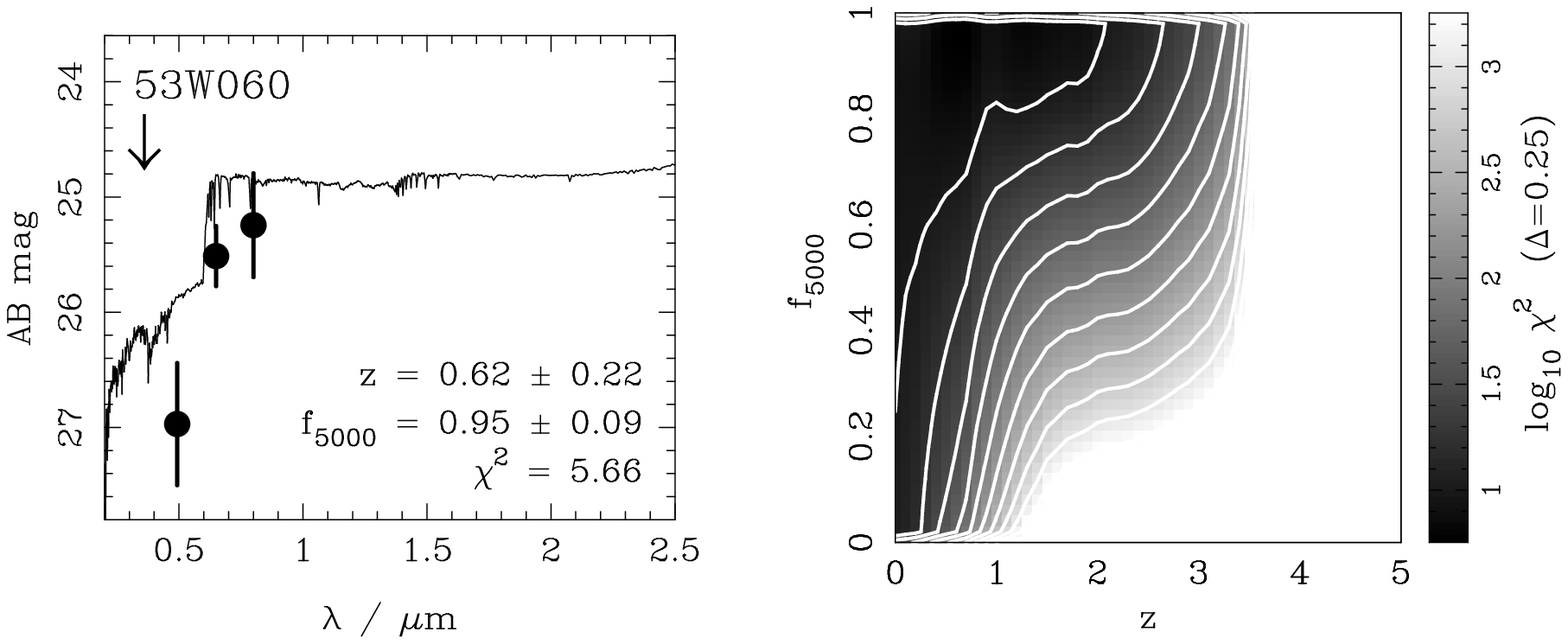,width=8cm}\ \ \ \ \ \ 
\psfig{file=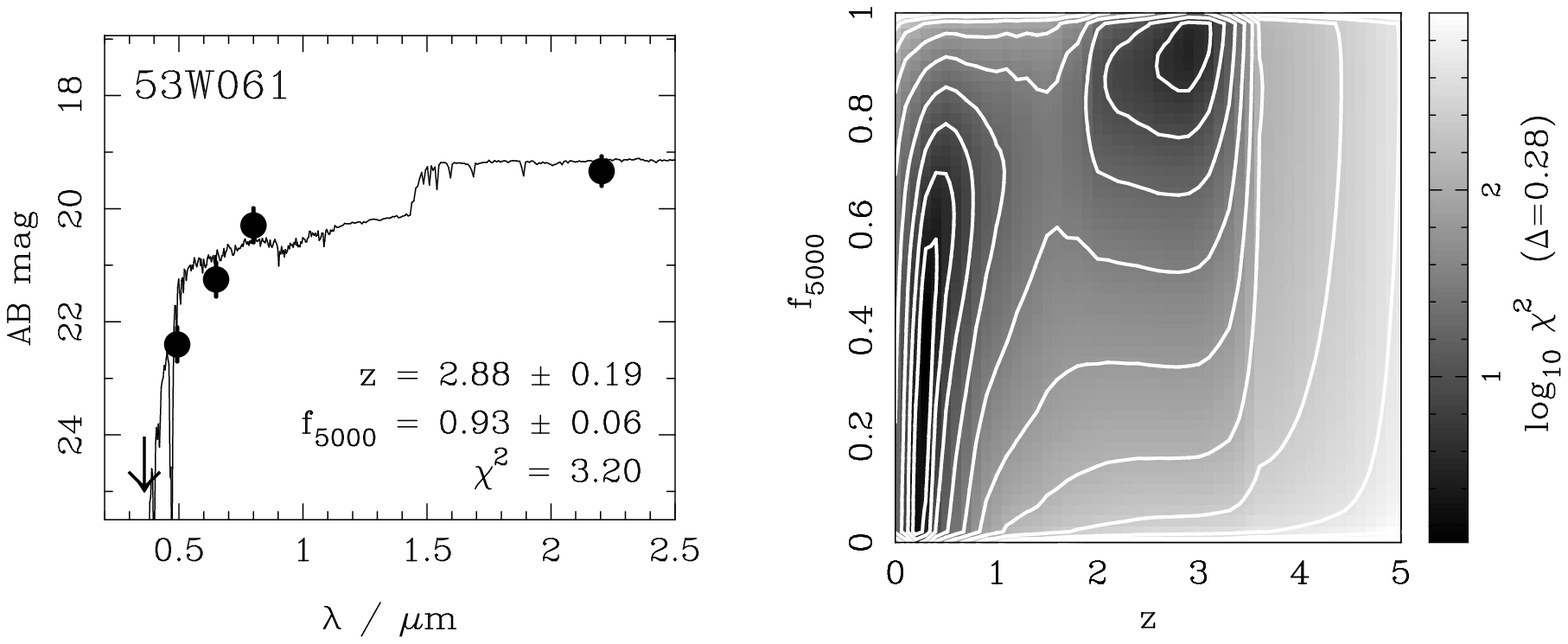,width=8cm}}
\hbox{}
\hbox{%
\psfig{file=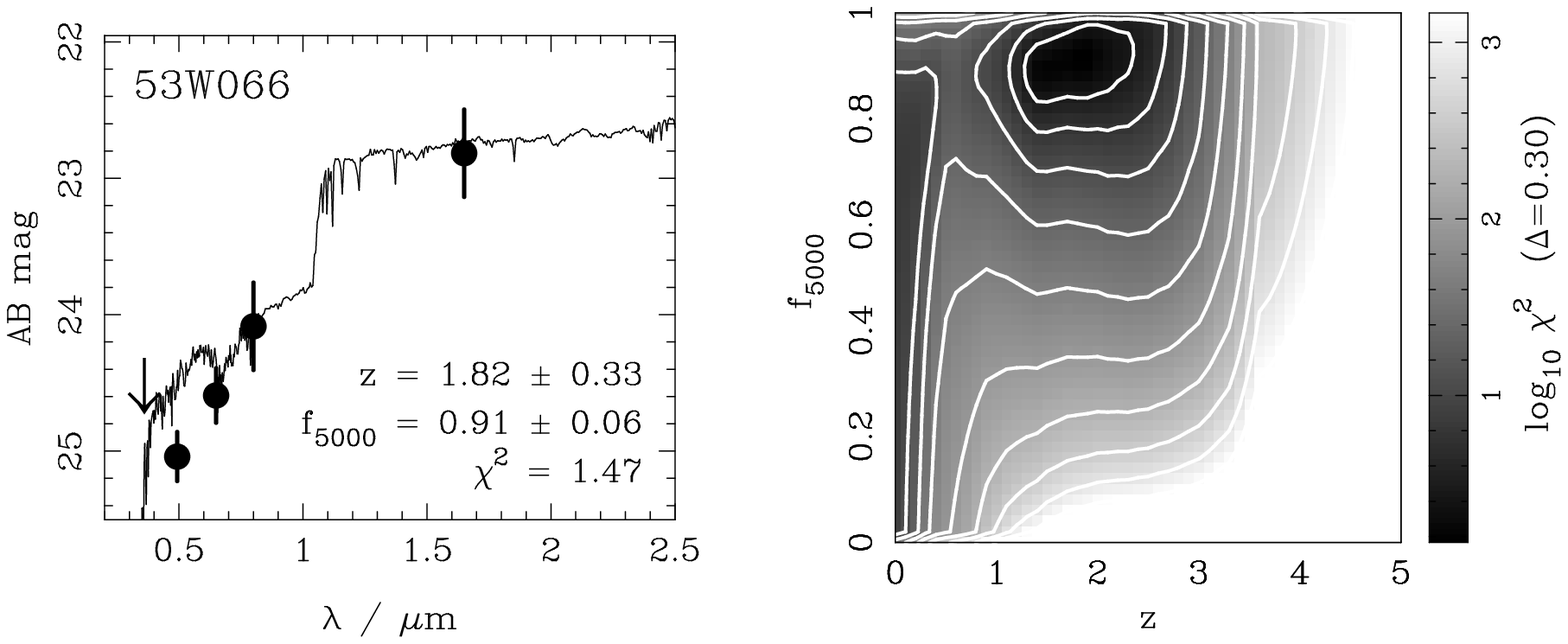,width=8cm}\ \ \ \ \ \ 
\psfig{file=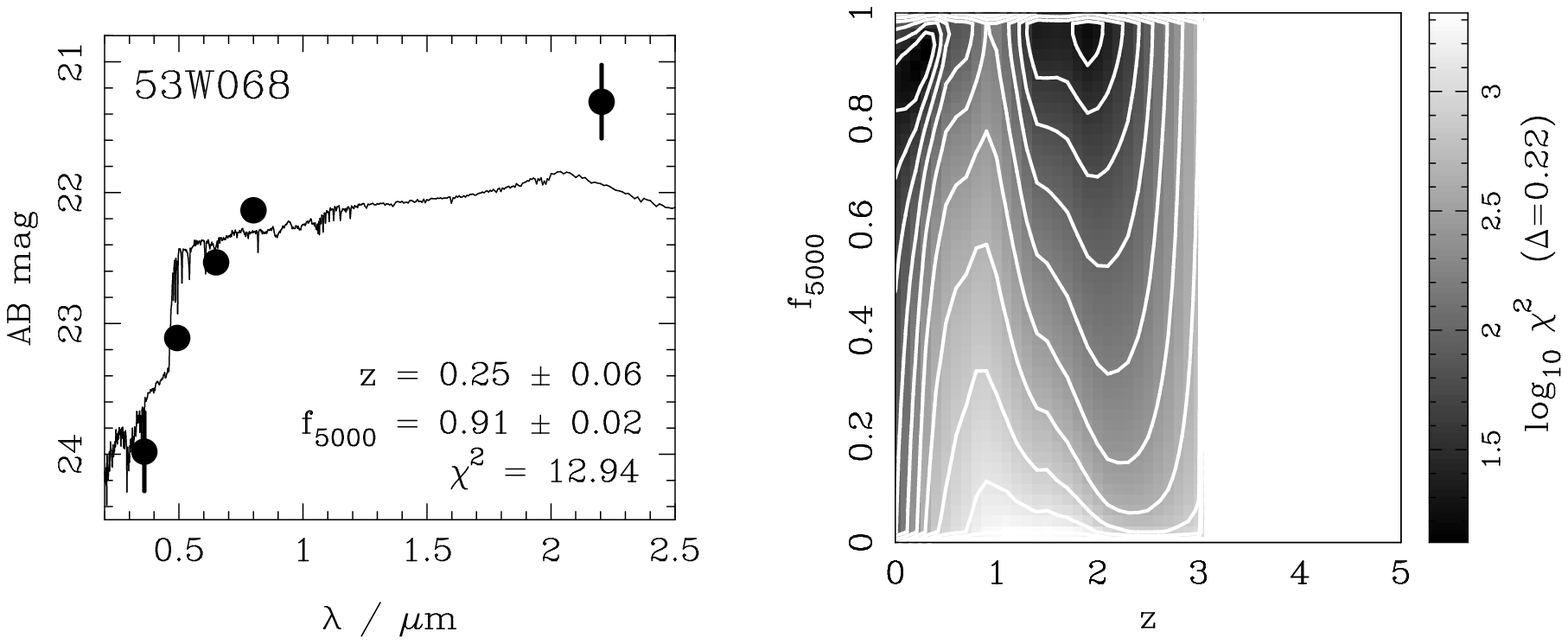,width=8cm}}
\hbox{}
\hbox{%
\psfig{file=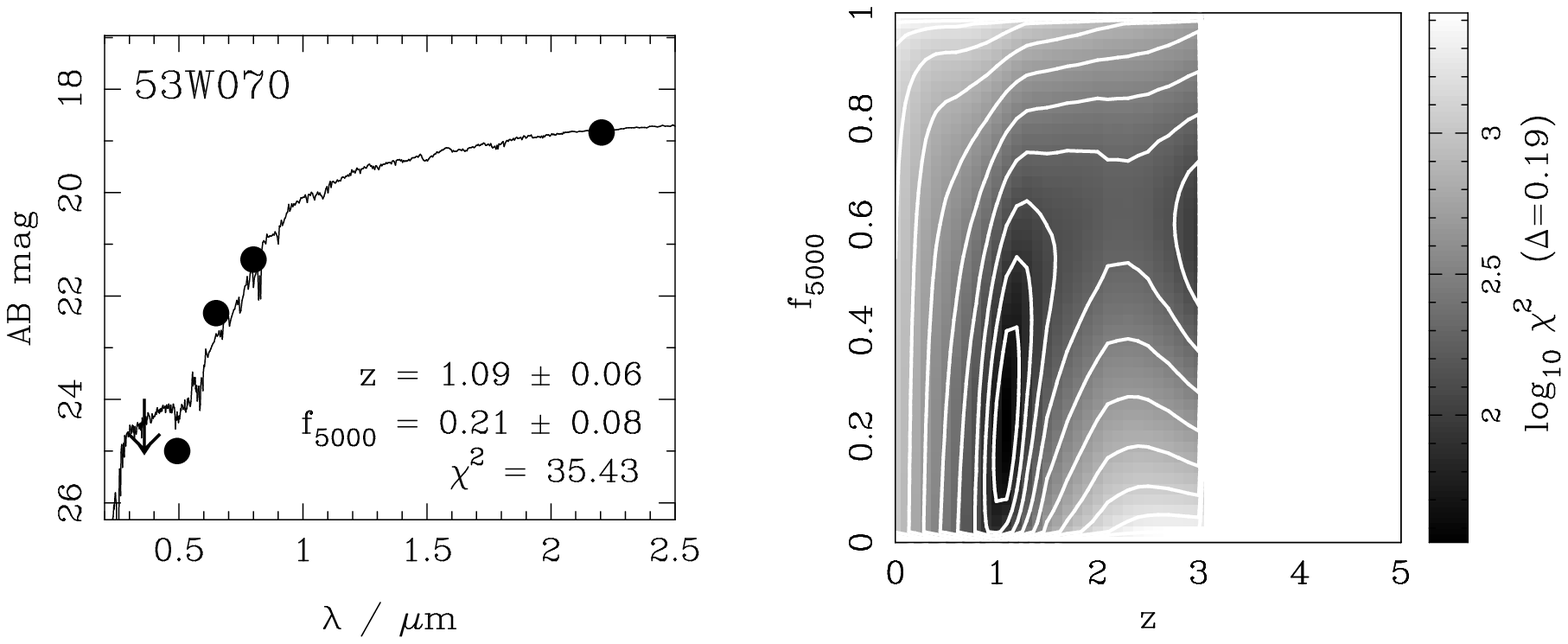,width=8cm}\ \ \ \ \ \ 
\psfig{file=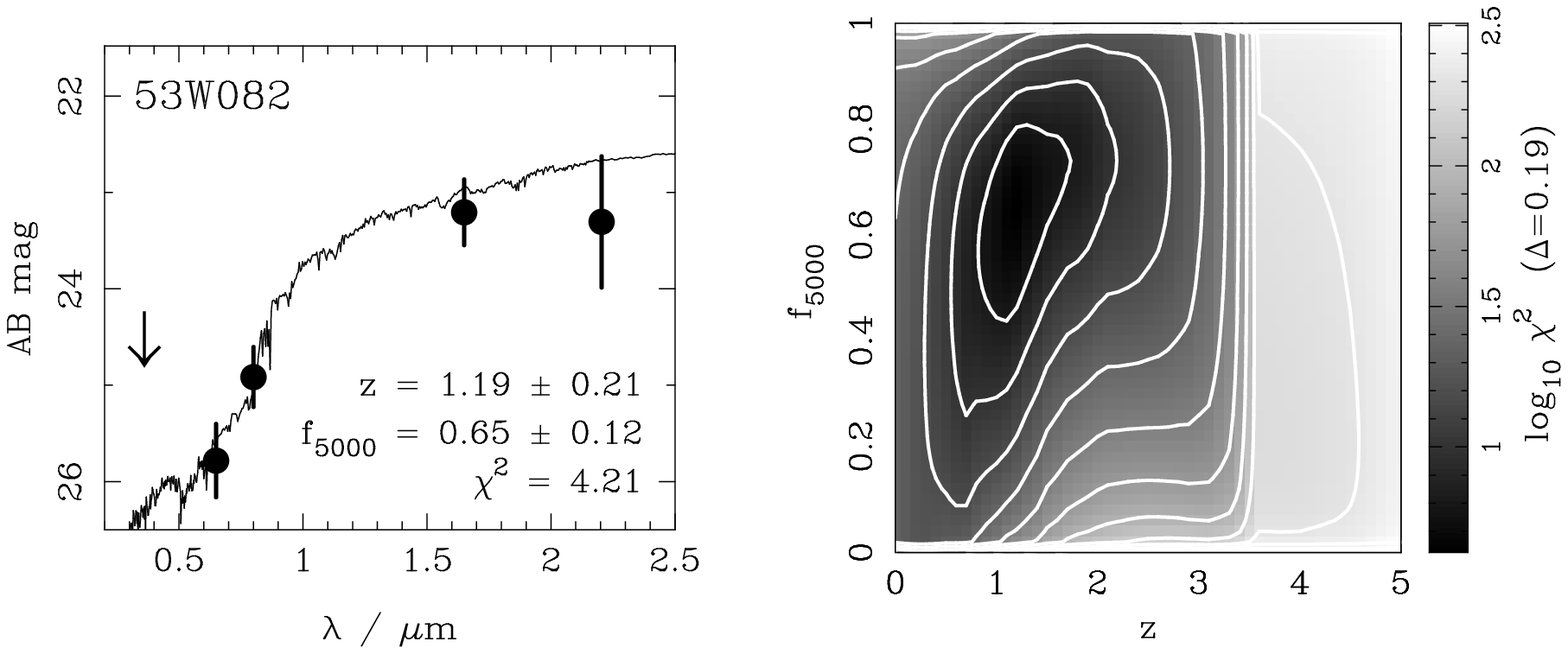,width=8cm}}
\contcaption{}
\end{minipage}
\end{figure*}


\begin{figure}
\psfig{file=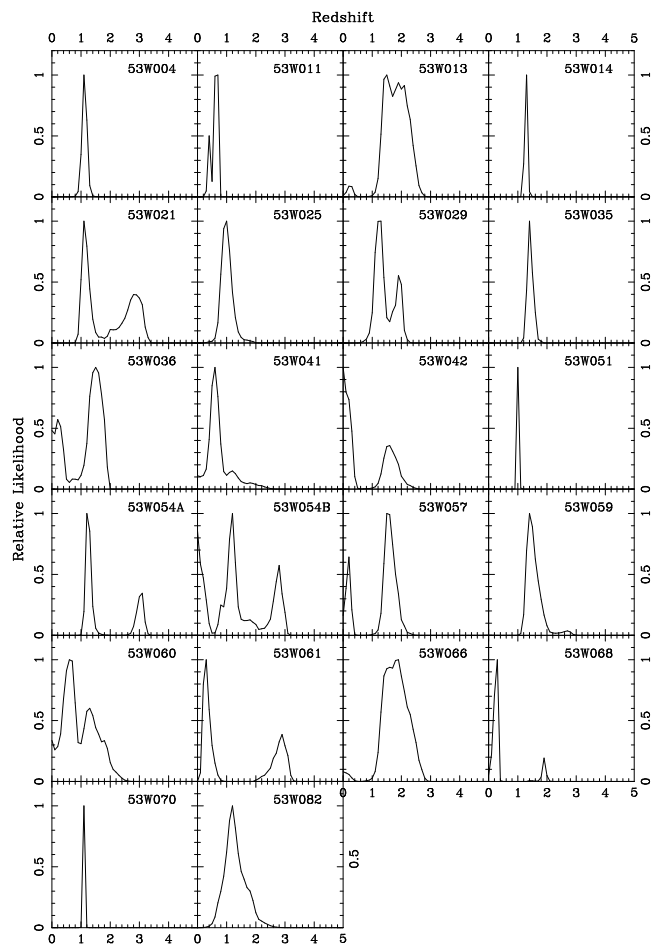,width=85mm}
\caption{Likelihood functions for the photometric redshifts of the
twenty-two sources in the LBDS Hercules sample without spectroscopic
redshifts.  $L(z)$ has been normalized to a peak value of
unity.\label{herclikelihood}}
\end{figure}


\begin{table}
\caption{Photometric redshifts for those twenty-two sources without
spectroscopic redshifts.  The one-sigma error on $z_{\rm phot}$ is
$\sigma_z$; \flilly\ is the fraction of the blue component in the
spectrum and $\sigma_f$ is its error.\label{photztable}}
\begin{tabular}{lccccc} \hline
Source  & $z_{\rm phot}$ & $\sigma_z$ & \flilly\ & $\sigma_f$ &
\chisq\ \\ \hline
53W004  & 1.12 & 0.09 & 0.64 & 0.05 &  1.23 \\
53W011  & 0.61 & 0.03 & 0.76 & 0.02 &  8.58 \\
53W013  & 1.49 & 0.22 & 0.90 & 0.04 &  7.82 \\
53W014  & 1.28 & 0.05 & 0.73 & 0.04 &  7.74 \\
53W021  & 1.12 & 0.11 & 0.15 & 0.16 &  6.53 \\
53W025  & 0.97 & 0.17 & 0.68 & 0.13 &  0.59 \\
53W029  & 1.23 & 0.15 & 0.43 & 0.09 &  5.49 \\
53W035  & 1.41 & 0.10 & 0.80 & 0.03 & 11.27 \\
53W036  & 1.50 & 0.28 & 0.98 & 0.01 &  0.70 \\
53W041  & 0.59 & 0.14 & 0.96 & 0.01 &  5.90 \\
53W042  & 1.58 & 0.25 & 0.97 & 0.01 &  4.78 \\
53W051  & 1.01 & 0.04 & 0.20 & 0.04 & 19.12 \\
53W054A & 1.25 & 0.08 & 0.51 & 0.06 &  3.94 \\
53W054B & 1.18 & 0.13 & 0.91 & 0.04 &  3.33 \\
53W057  & 1.53 & 0.17 & 0.97 & 0.01 & 14.95 \\
53W059  & 1.42 & 0.13 & 0.62 & 0.05 &  0.73 \\
53W060  & 0.62 & 0.22 & 0.95 & 0.09 &  5.66 \\
53W061  & 2.88 & 0.19 & 0.93 & 0.06 &  3.20 \\
53W066  & 1.82 & 0.33 & 0.91 & 0.06 &  1.47 \\
53W068  & 0.25 & 0.06 & 0.91 & 0.02 & 12.94 \\
53W070  & 1.09 & 0.06 & 0.21 & 0.08 & 35.43 \\
53W082  & 1.19 & 0.21 & 0.65 & 0.12 &  4.21 \\ \hline
\end{tabular}
\end{table}

For those sources which have an upper limit to the redshift based on a
continuum detection in the spectrum (table~\ref{speclimits}), this
limit was incorporated into the determination of $z_{\rm phot}$ and
the errors.  In fact, the spectroscopic continuum limit modified the
estimated redshift of only one source, 53W070, ruling out a high-$z$
minimum in \chisq\ at $z_{\rm phot} \simeq 3.7$.  

It can be seen in figure~\ref{herclikelihood} that several of the
sources have two or more distinct peaks in their redshift likelihood
function.  We generally adopted the most probable redshift (highest
peak) unless the alternative redshift was more consistent with the
$K$--$z$ or $r$--$z$ relation for the Hercules spectroscopic sample.
In this way we have chosen the second highest peak in $L(z)$ to be the
correct photometric redshift for two sources: 53W042 and 53W061.  With
an $r$-band magnitude of 25.7~mag (Paper I), the `peak' at zero
redshift for 53W042 cannot possibly correspond to the true redshift of
this source; the alternative of $z=1.58$ is prefered.  53W061 was
identified as a probable quasar \cite{Windhorst84b} and with $K=17.4$
the alternative redshift of 2.88 (with $\chi^2=3.2$) seems more
reasonable than $z=0.29$ (with $\chi^2=1.3$).


\begin{figure}
\psfig{file=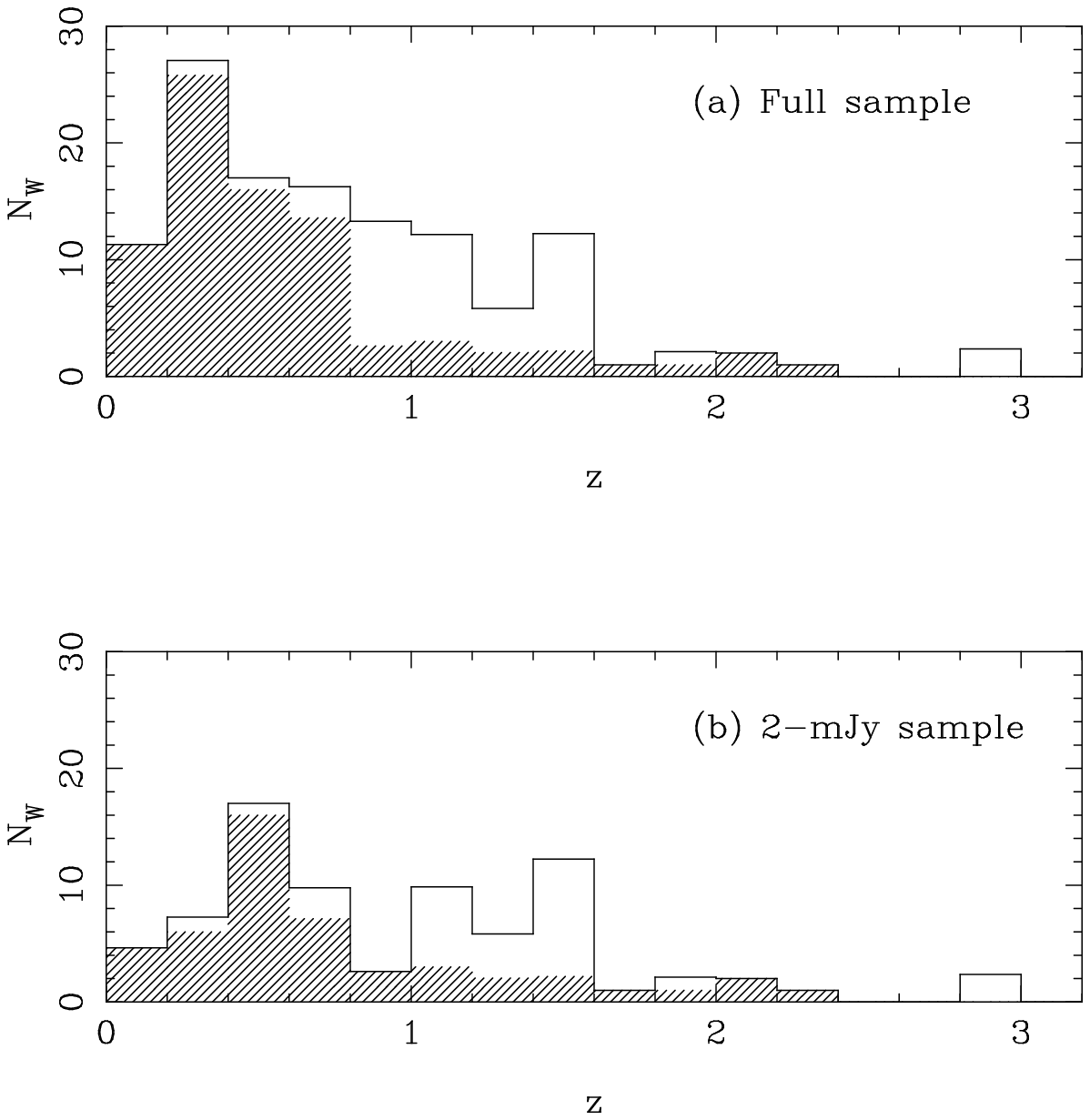,width=85mm}
\caption{Redshift distributions for the LBDS Hercules sample: (a) all
the sources; and (b) those sources with $S_{1.4}\ge 2$~mJy.  Hatched
histograms are the sources with spectroscopic redshifts, outlined
histograms are the sources with photometric redshifts.  Note that the
ordinate is the weighted number of sources -- i.e.\ corrected for
primary-beam attenuation and resolution bias.\label{herczhistograms}}
\end{figure}

The redshift distribution of the LBDS Hercules sources is presented in
figure~\ref{herczhistograms}, for both the full sample and the 2-mJy
sample.  In this figure the photometric redshifts from
table~\ref{photztable} have been combined with the spectroscopic
redshifts from table~\ref{linelist} here and table~3 of Paper~I.  The
numbers have been appropriately weighted to correct for the detection
biases in the radio data (see Paper~I).  The median redshift of the
full sample is 0.61, and for the 2-mJy sample it is 0.80.  Comparing
these results with fainter radio surveys at sub-millijansky and
microjansky flux density limits, it is seen that the redshift
distribution changes very little over more than two orders of
magnitude in radio flux \cite{Windhorst98}.  In particular, radio
observations of the Hubble Deep Field down to 9~$\mu$Jy result in a
very similar distribution to figure~\ref{herczhistograms}(b), with a
mean redshift of $\sim 0.8$ \cite{Richards98}.

Recall that there are three sources in the Hercules sample that are
unidentified and therefore do not appear in the redshift distributions
(Paper~I).  53W043 was obscured by a bright star in the \fsh\ CCD
observations, but a limit of $F^+ > 23$~mag was obtained from the
photographic plates.  This places a lower limit on its redshift of
$z\ga 0.5$.  The other two unidentified sources (53W0037 \& 53W0087)
both have $r \ga 25$~mag, placing them at $z \ga 1$--2.  With a
combined weight of 3.7, these three sources will not significantly
change the distribution but will likely extend the high-redshift tail
when they are finally identified.  We calculate an upper limit to the
median redshift of the whole 72-source sample of 0.63 and an upper
limit of 1.09 for the 2-mJy sample, taking into account these
unidentified sources.


\begin{figure}
\psfig{file=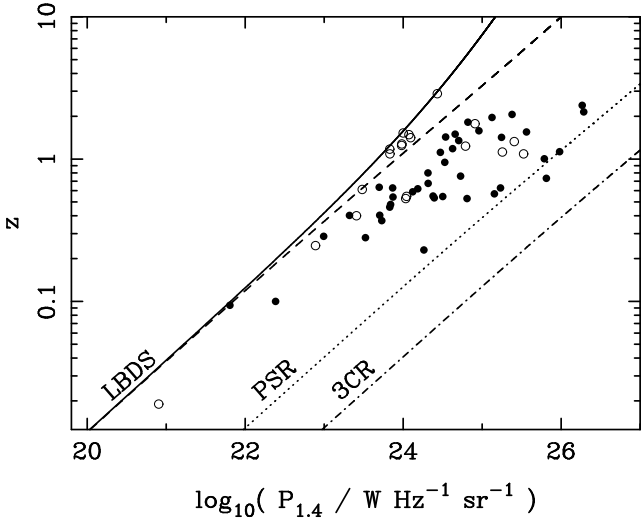,width=85mm}
\caption{The luminosity--redshift plane for sources with $S_{1.4}\ge
2$~mJy in the LBDS Hercules field.  The flux density limits for the
survey are shown for flat-spectrum (solid line, open circles) and
steep-spectrum (dashed line, solid circles) sources, together with the
limits for the PSR (dotted line) and 3CR (dot-dash line) surveys.
Photometric redshifts have been used for those sources without a
spectroscopic measurement.\label{fluxlimits}}
\end{figure}

In figure~\ref{fluxlimits} we compare the flux density limits of the
LBDS with those of the Parkes Selected Regions (PSR; Dunlop \etal\
1989)\nocite{Dunlop89} and 3CR \cite{Laing83,Spinrad85} surveys.  It
can be seen how the LBDS can be used to: (i) probe the faint end of
the RLF out to much greater redshifts than the brighter surveys; and
(ii) detect powerful radio galaxies out to very high redshifts ($z\ga
10$).  With a flux density limit of $S_{1.4} = 2$~mJy, the LBDS is
$\sim$100 times fainter than the PSR used by DP90.  In the PSR sample,
only the most powerful ($\logten P_{1.4} > 26$) sources are detectable
at $z>1.5$.  The LBDS can detect sources at the same redshift which
have luminosities of $\logten P_{1.4} \simeq 24$ -- such sources are
below the flux density limit of the PSR for any redshift $z>0.1$.  If
there is not a high-redshift cut-off in the RLF, then there will be a
significant fraction of sources in the LBDS at high redshifts.  For
example, figure~14(b) of DP90 predicts the cumulative redshift
distribution for the whole LBDS sample based on their RLF models.  As
many as 10\% of the sources may have redshifts greater than 3
according to the models, although the predicted distribution was
rather poorly constrained beyond $z\ga 0.7$.  This ability of the LBDS
to confirm or refute the cut-off was one of the motivations to
complete the optical identification of the LBDS and to obtain
redshifts for as many sources as possible.  The lack of sources at
$z\ga 2.5$ in figure~\ref{fluxlimits} is striking, and clearly
suggests that there is indeed a redshift cut-off in the RLF.  In the
next section, we quantify this statement and compare the data with the
models of DP90.

\section{The radio luminosity function and redshift cut-off}

\subsection{A review of the RLF models}

In an extension of work begun by Peacock \& Gull
(1981)\nocite{Peacock81} and Peacock (1985)\nocite{Peacock85}, DP90
investigated the evolution of the 2.7-GHz radio luminosity function,
with an emphasis on the behaviour of the RLF at high redshift.  Their
method was to find a model (or rather, an ensemble of models) of the
luminosity function \rlf, that was consistent with all the available
data.  In those regions of the $P$--$z$ plane where the redshift
content of the data was high, generally corresponding to the higher
flux densities, the model was well defined.  The model was then
extrapolated across the rest of the $P$--$z$ plane, subject to the
constraints of less direct data such as source counts.  Of course, it
is known that the real universe is not smooth on all scales, so the
models will only be approximations to the actual form of the RLF.

The data were drawn from four complete samples at a frequency of
2.7-GHz, together with source counts at fainter flux limits and
measurements of the local RLF.  The deepest of these samples was the
Parkes Selected Regions (Downes \etal\ 1986; Dunlop \etal\
1989)\nocite{Downes86,Dunlop89}, with a flux density limit of 0.1~Jy
and containing 178 sources (including a higher proportion of
steep-spectrum sources than the other samples).  However, only 46\% of
the PSR sources had redshifts, so DP90 used the empirical $K$--$z$
relation of Lilly, Longair \& Allington-Smith (1985)\nocite{Lilly85}
to estimate the redshifts of the remaining 54\% of the data.

The ensemble of model RLFs consisted of five `free-form' and two
parametric models that were found to be consistent with the data.  The
free-form models (here denoted by FF-1, $\ldots$, FF-5) are forth- or
fifth-order series expansions in $\logten P_{2.7}$ and $z$, with
various forms of high-redshift cut-off imposed on the models.  Each of
these models was derived twice, using two different estimated redshift
distributions (denoted by `MEAN-z' and `HIGH-z').  DP90 also fitted
their data with two simple parametric models: pure luminosity
evolution (PLE) and a combination of luminosity and density evolution
(LDE).  The PLE luminosity function was modelled as the sum of a
low-power non-evolving component (a sixth-order polynomial in $\logten
P_{2.7}$) and a high-power evolving component (a double power-law in
$P_{2.7}$, with a break luminosity parameterized as a quadratic
function of $z$).  The LDE model was a modification of PLE, in which
the space density was allowed to vary with redshift directly, in order
to investigate the possibility that the cut-off was due to negative
density evolution while the positive luminosity evolution continued at
$z\ga 2$.

For the flat-spectrum population, a decline in the RLF at high
redshifts ($z\ga 2$) is required by the models for all luminosities
where the complete sample database has good coverage of the $P$--$z$
plane (the strength of this result is due to the nearly complete
redshift data for the flat-spectrum population).  For the
steep-spectrum sources, a cut-off is {\it required\/} for the most
luminous objects ($\logten P_{2.7}\ge 27$) at $z\ga 2$, and it is {\it
consistent\/} with the data at all luminosities.  This was the first
time that a high-$z$ cut-off in the RLF had been quantified in the
steep-spectrum population, although it had also been suggested by the
observations of Windhorst (1984)\nocite{Windhorst84}.

DP90 also performed a model-independent test of the cut-off using the
banded \vvmax\ test \cite{Schmidt68,Rowan-Robinson68,Avni83}.  For a
uniform distribution of objects in space, the mean $\langle V/V_{\rm
max} \rangle$ should be 0.5.  The banded \vvmax\ test is a function of
redshift and enables any high-redshift negative evolution to be
separated from the strong, positive evolution at lower redshifts.
Applying this test to their complete sample database gave values
$\langle V/V_{\rm max} \rangle < 0.5$, on average, for the
flat-spectrum sources at $z\ga 1$ and for the steep-spectrum sources
at $z\ga 1.5$, indicating a deficit of high-redshift objects.  At
lower redshifts, values of $\langle V/V_{\rm max} \rangle > 0.5$
clearly demonstrated the strong positive evolution of both the steep-
and flat-spectrum populations at $z \la 1$.

\subsection{The cumulative redshift distribution}

The models of DP90 were used to extrapolate the RLF to lower radio
luminosities and higher redshifts than were sampled by their data.  In
fact the models predict the form of the RLF over the whole of the
$P$--$z$ plane illustrated in figure~\ref{fluxlimits}.  Thus we can
use the LBDS Hercules sample to test the reality of the redshift
cut-off found by DP90.  As we discussed in Paper I, the large weights
of the nine sources with $1\le S_{1.4} < 2$~mJy causes the full sample
to be biased towards these low signal-to-noise objects.  Throughout
this section, therefore, only the 63 sources with $S_{1.4} \ge 2$~mJy
(i.e.\ the `2-mJy sample') will be considered.

Two approaches have been used in order to compare the predictions with
the Hercules data.  First, the model luminosity functions were used to
calculate the expected redshift distribution in the Hercules field, in
the form of cumulative number counts.  Second, the space density of
sources in the sample was calculated as a function of radio luminosity
and redshift, to compare directly with the models.  This second method
has the advantage that the luminosity dependence can be investigated
directly, but with only 63 sources in the sample (of which three have
no optical counterpart) the sampling and statistics are quite poor.
The first method allows more robust conclusions to be drawn, but at
the expense of losing the luminosity information.

DP90\nocite{Dunlop90} divided their data according to spectral index
($\alpha$, where $S_\nu \propto \nu^{-\alpha}$), and calculated model
RLFs for the steep-spectrum ($\alpha \ge 0.5$) and flat-spectrum
($\alpha < 0.5$) populations separately.  For their complete samples,
spectral indices were known for all sources.  However, the data from
the fainter samples were generally restricted to source counts, with
no spectral index information available.  To separate the faint number
counts into steep- and flat-spectrum sources, they approximated the
flat-spectrum contribution by a model, ${\rm d}n_{\rm flat}/{\rm d}S =
45\, S^{-2.5}\, \exp [ - (\ln S)^2 / 8 ]$~Jy$^{-1}$~sr$^{-1}$, which
was then subtracted off the data to give the steep-spectrum counts.
These number counts were incorporated into the respective RLF models
of the two populations.

It became apparent during the current work that this model for the
flat-spectrum contribution at faint flux densities was incorrect.  The
number of steep-spectrum sources in the Hercules data is significantly
lower than the RLF models predicted, and conversely there are more
flat-spectrum sources in the data than in the models.  The predicted
total number of sources (steep- plus flat-spectrum) agreed with the
data, as it should.  In particular, DP90's model for ${\rm d}n_{\rm
flat}/{\rm d}S$ predicts that there should be 5--10 flat-spectrum
sources in the 2-mJy sample, where there are actually 29 of them
(after applying the weights for incompleteness).  In fact, the ratio
of steep-spectrum to flat-spectrum sources is the same in both the PSR
and 2-mJy Hercules samples, at approximately 2$\,$:$\,$1 (see also
Windhorst \etal\ 1993; Richards 2000).\nocite{Windhorst93,Richards00}

We therefore removed the distinction between flat-spectrum and
steep-spectrum sources altogether.  The data were considered as a
single population, and the models were simply added together.
Mathematically, this is possible as the RLF models are essentially the
number of sources observed in the survey, which can of course be
summed.  Physically, this is justified by the work of several authors.
One of the main conclusions of DP90 was that both populations behave
very similarly and may therefore come from a single population.
Padovani \& Urry (1992)\nocite{Padovani92} demonstrated that steep-
and flat-spectrum radio quasars and FR-II radio galaxies were
consistent with being the same physical objects, viewed at different
angles and with a range of relativistic beaming parameters (see also
Urry \& Padovani~1995).\nocite{Urry95} Thus in the remainder of this
paper we consider a single radio luminosity function for sources of
all spectral indices.

\begin{figure}
\psfig{file=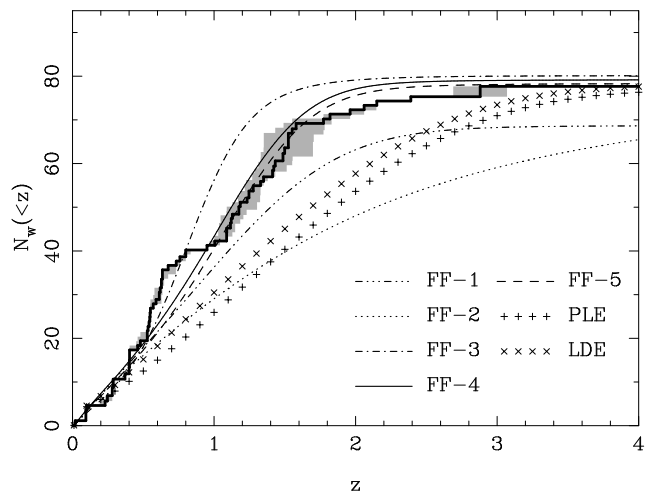,width=\hsize}
\caption{The cumulative redshift distribution of sources in the 2-mJy
Hercules sample.  The bold histogram is computed from the combined
spectroscopic and photometric redshift distribution of the sample.
The shaded area is a measure of the uncertainty in the distribution,
assuming the photometric redshifts were all at their $\pm1$-$\sigma$
errors. The model radio luminosity functions of DP90 are also plotted:
free-form models 1 through 5 (FF-1, $\ldots$, FF-5), pure luminosity
evolution (PLE) and luminosity/density evolution (LDE).  $N_W$ is the
weighted number of sources with redshifts less than $z$.\label{cumz1}}
\end{figure}

The cumulative number counts are calculated for each of the RLFs: five
free-form models, PLE and LDE.  The total number of sources at
redshifts less than $z$ is:
\begin{equation}
N(<z) = \int_0^z \left[ \int_{P_1}^{\infty} \rho(P,z) \
        {\rm d}P \right] \left( {{{\rm d}V}\over{{\rm d}z}}
        \right) {\rm d}z
\label{nlessthanz}
\end{equation}
where $P_1$ is the luminosity of a source at redshift $z$ whose flux
density would be equal to the detection limit of $S_{2.7}=1.2$~mJy.
This projected flux density limit at 2.7~GHz is derived from the flux
density limit of the survey at 1.4~GHz (i.e.\ 2.0~mJy) using the
median spectral index of sources in the sample, $\alpha_{\rm med} =
0.77$, defined over 0.6--1.4~GHz (Paper~I).  Equation~\ref{nlessthanz}
was evaluated numerically in steps of $\Delta \logten P = 0.025$ and
$\Delta z = 0.01$.  The volume element is ${\rm d}V \simeq \Delta V =
[ V(z+\Delta z) - V(z) ] \Psi$, where the comoving volume of the
universe (in Mpc$^3$~sr$^{-1}$) out to redshift $z$ is $V(z)$, and
$\Psi=3.78\times10^{-4}$~sr (or 1.22~deg$^2$) is the solid angle of
the sample \cite{Windhorst84b}.  The results are plotted in
figure~\ref{cumz1}.  The observed cumulative redshift distribution was
computed by summing the weights of the sources from $z=0$ to the
redshift of each source in turn.  We made a conservative calculation
of the uncertainty in this distribution by assuming that all the
photometric redshifts were simultaneously at their $\pm1$-$\sigma$
errors (shaded area in figure~\ref{cumz1}).


Five of the models correctly predict the total number of sources in
the sample, to within 5\%.  The weighted number of sources is 81.4, of
which 3.7 are unidentified and do not appear in the figure, and the
models predict $N_{\rm total} \simeq 78$.  The other two models (FF-1
and FF-2) converge to $N_{\rm total} \simeq 73$ by $z=8$.  

The best agreement with the data is obtained for free-form models FF-4
and FF-5.  At $z<0.5$ and $z>1$ they follow the observed distribution
closely.  The only disagreement occurs over the range $0.5 \la z \la
1$, where the observed redshift distribution rises more rapidly than
the models.  This rise is due to two peaks in the observed redshift
distribution -- there are seven sources at $z = 0.54 \pm 0.02$ and
five sources at $z = 0.62\pm 0.01$ in the 2-mJy sample (plus another
source at $z = 0.61$ with $S_{1.4}=1.7$~mJy).  These two peaks
indicate the existence of large-scale structures (sheets) intersecting
the line-of-sight.  We suggest that these radio sources may be part of
two super-clusters (their comoving transverse separation is
$\sim25$~Mpc), statistical variations that the RLF models could not
possibly predict.

With these exceptions, models FF-4 and FF-5 provide a good description
of the LBDS Hercules data.  Recall that the difference between each of
the free-form models lies in the form of the high-redshift cut-off.
Although these two models are very similar in their predictions, they
were defined somewhat differently.  A high-$z$ cut-off was enforced on
FF-5 such that the RLF decays sinusoidally from $z=2$ to a value of
zero at $z>5$.  For FF-4, the model was similarly constrained to a
value of zero at $z>5$, but the form of the cut-off was not specified
(beyond being smooth and continuous).

The other five models all fit the observed distribution at low
redshifts, $z\la 0.4$, where they were relatively well-constrained by
the local RLF, but at high-$z$ they fail to match the data
(figure~\ref{cumz1}).  The second free-form (FF-2), pure luminosity
evolution (PLE) and luminosity/density evolution (LDE) models in
particular, predicted that a much greater proportion of the sources
would have large redshifts.  Using a preliminary version of this
sample, Dunlop (1997)\nocite{Dunlop96b} found that the PLE and LDE
models were consistent with the high-redshift ($z>2$) distribution of
the data.  However with the additional spectroscopic and photometric
redshifts now available, we see that these models show a large
disagreement with the data over lower redshifts, $0.5\la z\la 2.5$.
This is primarily due to the fact that the improved SED-based redshift
estimation procedure used here has proven that the redshifts for many
sources were previously over-estimated on the basis of the $K-z$
relation for powerful radio galaxies.

The first of the free-form models (FF-1) approximately follows the
redshift distribution from $z=0$ to $z\sim1$, however the number of
sources at higher redshifts is $\sim$15\% too low.  The third of the
free-form models (FF-3) is in reasonable agreement with the data for
redshifts $z \la 0.8$, but then continues to rise steeply.  It
predicts that there should be only five sources at $z>1.5$, which is
inconsistent with the data as there are already ten sources with
spectroscopic redshifts greater than this.

\subsection{The binned radio luminosity function}

\begin{figure}
\psfig{file=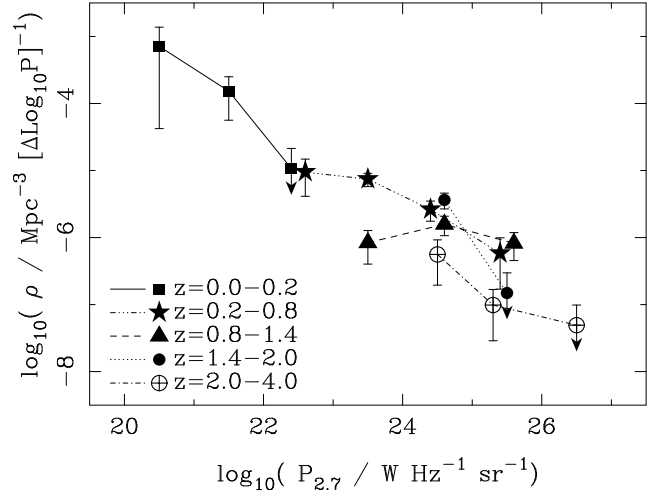,width=\hsize}
\caption{The observed radio luminosity function for the 2-mJy Hercules
sample.  The different symbols and lines correspond to different
redshifts, as shown, with arrows representing bins which contain a
single source.  The RLF has been plotted against the 2.7~GHz
luminosity of each source (extrapolated from the 1.4~GHz luminosity)
to aid comparison with the models.\label{rlfdata}}
\end{figure}

Figure~\ref{cumz1} has shown which models correctly predict the
overall redshift dependence of the radio luminosity function, but it
says nothing about the luminosity dependence.  To investigate this, we
calculated the observed RLF to compare directly with the models.  This
was done by redshift and luminosity binning of the 2-mJy Hercules
data.  The luminosity of each source at 2.7~GHz (the frequency at
which the RLF models are defined) was calculated from the observed
flux at 1.4~GHz and the 0.6--1.4~GHz spectral index.  Eight luminosity
bins were used, from $\logten P_{2.7} = 20$ to 28 in steps of $\Delta
\logten P_{2.7} = 1$.  Five redshift bins were used, from $z=0$ to 4,
centred on $z=0.1$, 0.5, 1.1, 1.7 and 3.0.  These redshifts
approximately correspond to ages of 90\%, 50\%, 33\%, 25\% and 10\% of
the current age of the universe (for an Einstein-de~Sitter cosmology
where $(1+z) \propto t^{-2/3}$).  The bins were chosen so as to
contain enough sources to be statistically useful, while being
sufficiently narrow to resolve the redshift dependence of the RLF.
The binned luminosity function is presented in figure~\ref{rlfdata}.
While the small number of sources is the weakness of this method of
analysing the data, its strength lies in the accuracy with which the
luminosities are known, given the effectively complete redshift
information for the sample.


Given the relatively large width of the redshift bins, an additional
weight had to be applied to each source to correct for the fact that
the weakest sources could not have been detected at all redshifts
within the bin.  Specifically, each weight was scaled by $V / V_{\rm
lim}$, where $V = V(z_2) - V(z_1)$ is the comoving volume enclosed by
the bin extending from redshift $z_1$ to $z_2$, and $V_{\rm lim} =
V(z_{\rm lim}) - V(z_1)$ is the volume enclosed from $z_1$ to the
redshift $z_{\rm lim}$ at which the source would fall below the radio
flux density limit of the sample (clearly if $z_{lim} \ge z_2$ then
this weighting was unnecessary).  Errors were assigned to each bin on
the basis of Poisson counting statistics, applied to the raw weighted
counts before correcting for the flux-density-limit bias.

It is seen in figure~\ref{rlfdata} that the observed low-redshift RLF
evolves more slowly at low luminosities than it does at the higher
luminosities studied by DP90.  It is not until $z\ga 0.5$ that we see
significant evolution in the luminosity bins.  At $\logten P_{2.7} =
23.5$, the density of sources at $z=1.1$ (triangles in
figure~\ref{rlfdata}) is reduced by an order of magnitude from that in
the lower redshift bin at $z=0.5$ (stars).  Note that this may be due
to incompleteness at the survey limit, but that is unlikely -- we
showed in Paper~I that the sample is effectively complete once the
appropriate weights have been applied to the sources.  There is no
evidence for a declining RLF at the same redshift ($z=1.1$) in the
next luminosity bin ($\logten P_{2.7} = 24.5$), suggesting that the
cut-off redshift is a function of the radio luminosity.

One of the most significant features in figure~\ref{rlfdata} is the
behaviour of the high-redshift radio luminosity function (encircled
crosses).  All three points defining the RLF at $z>2$ lie below those
of the same luminosities ($\logten P_{2.7} = 24$--27) at $z<2$.  In
particular, the density of sources with $\logten P_{2.7} = 24.5$
decreases by a factor of six between $z=1.7$ and 3.0.  The decline at
a luminosity of $\logten P_{2.7} = 25.5$ appears to be less dramatic
and is of low formal significance due to there being just a single
source at $z=1.7$ at this luminosity.  It can be seen that the density
has dropped by a factor of ten between $z=1.1$ and 3.0 over this same
luminosity interval.  We conclude from our direct binning of the RLF
that the comoving density of low-luminosity radio sources begins to
decline at redshifts $1\la z\la 2$ and that the actual redshift of the
turnover may depend on the radio luminosity.


\begin{figure}
\psfig{file=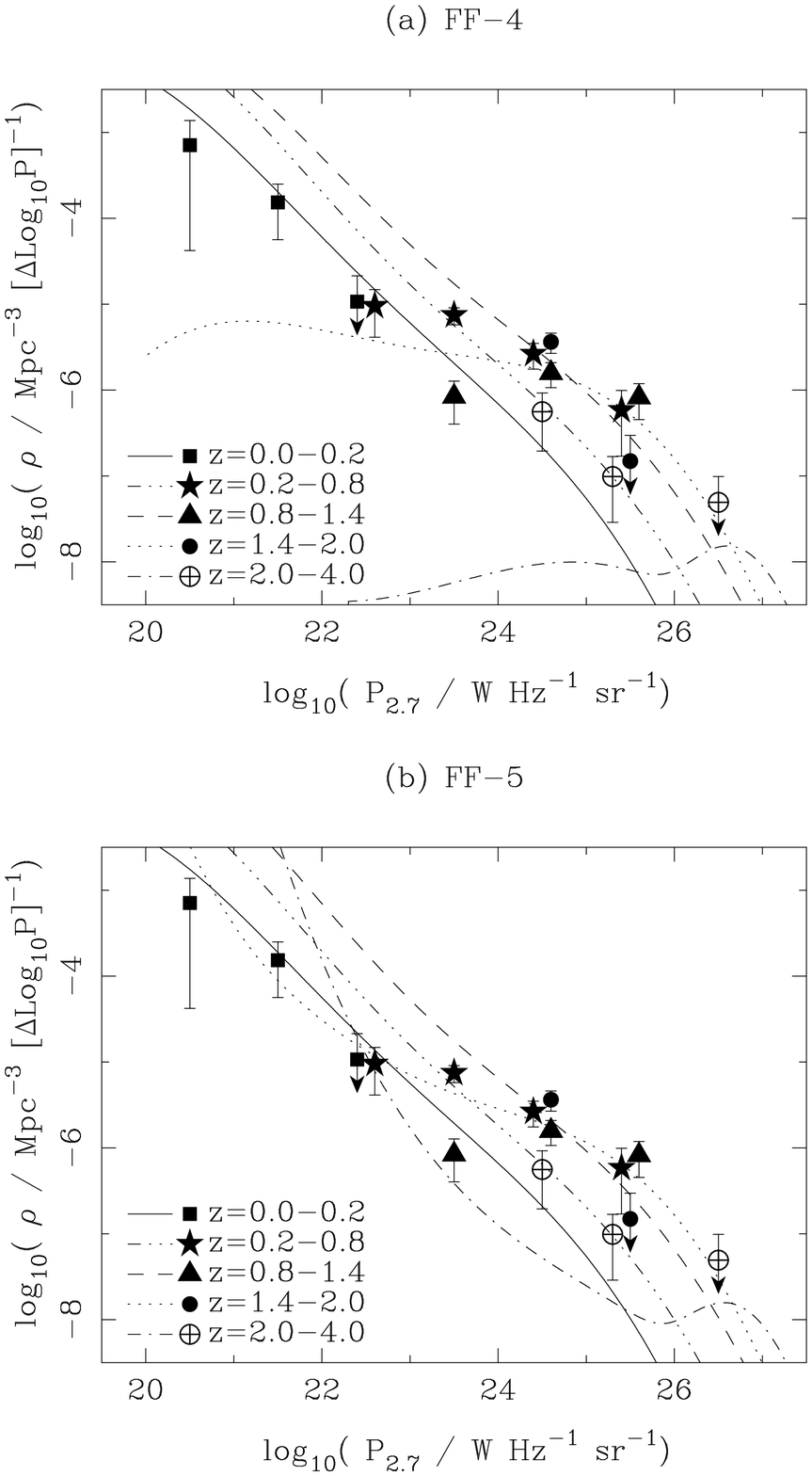,width=\hsize}
\caption{The observed radio luminosity function (points) for the 2-mJy
Hercules sample compared with the two model RLFs of DP90 which
correctly predicted the redshift distribution of the sample: (a) FF-4,
and (b) FF-5.\label{rlfcompare}}
\end{figure}

In figure~\ref{rlfcompare} we compare the observed luminosity function
with the two model RLFs of DP90 which correctly predicted the redshift
distribution of the sample, i.e.\ free-form models FF-4 \& FF-5.  At
the lowest redshifts, $z<0.2$ (squares, solid lines in
figure~\ref{rlfcompare}), the model RLFs are consistent with the data
within their errors.  At a redshift of $z\sim 0.5$ (stars,
dash-dot-dot-dot lines) both models predict too many sources at
$\logten P_{2.7} \simeq 22$--23 and too few at $\logten P_{2.7} \simeq
24$--26, although the data differ from the models by only a few sigma.
At $z\sim 1.1$ (triangles, dashed lines) the data and models are in
reasonable agreement at $\logten P_{2.7} > 24$, but there are
significantly fewer low luminosity ($\logten P_{2.7} \simeq 23$--24)
sources than predicted.

It is the behaviour of the RLF at the highest redshifts that is
crucial for measuring the redshift cut-off.  Both the free-form models
FF-4 and FF-5 are consistent with the data at $z\sim 1.7$ (filled
circles, dotted lines in figure~\ref{rlfcompare}), within their
two-sigma error bounds.  It is seen that at this redshift the model
RLFs begin to decline for sources with luminosities $\logten P_{2.7} <
25$, although the observed RLF does not change significantly from
$z=0.5$ to 1.7 at these luminosities.  The redshift cut-off of DP90 is
clearly seen in the highest-redshift bin in figure~\ref{rlfcompare}(b)
and is even more prominent in figure~\ref{rlfcompare}(a).  The
comoving density of radio sources at $z\sim 3$ is predicted to decline
by as much as two orders of magnitude from $z\sim 1.7$.  However, the
data do not fall off nearly as rapidly as the models.  Model FF-5 is
closer to the observed RLF than model FF-4 in this high-redshift
regime, and is consistent with the data at the two-sigma level.

We noted above that with an average of four sources per binned data
point in figures~\ref{rlfdata} \& \ref{rlfcompare}, the statistics are
quite poor and the data do not differentiate between the models to a
significant degree.  However, we have shown that the two models which
best fit the observed redshift distribution of the sample are at least
consistent with the luminosity dependence of the observed RLF.

\subsection{The banded V/V$_{\rm max}$ test}

Following DP90, we applied the banded \vvmax\ test to the Hercules
sample in order to investigate the redshift cut-off in a model
independent manner.  This is a variation of the original \vvmax\ test
\cite{Schmidt68,Rowan-Robinson68}, that has been modified to test the
evolution of the sample for redshifts greater than some fixed value
$z_0$.  We define $V$ as the comoving volume enclosed by a source at
redshift $z$ ($\ge z_0$), $V_{\rm max}$ as the volume that would be
enclosed by the same object if it were pushed out to the flux density
limit of the sample, and $V_0$ as the volume enclosed within a
redshift of $z_0$.  If a population of $N$ sources is uniformly
distributed for redshifts $z \ge z_0$, then $(V-V_0)/(V_{\rm
max}-V_0)$ would be distributed uniformly between 0 and 1, with a mean
of $0.5\pm (12N)^{-0.5}$ \cite{Avni80,Avni83}.  Values of the mean
$\langle (V-V_0)/(V_{\rm max}-V_0) \rangle > 0.5$ correspond to an
increase in comoving density with redshift, while $\langle
(V-V_0)/(V_{\rm max}-V_0) \rangle < 0.5$ indicates a deficit of
high-redshift objects.  Note that the values of $\langle
(V-V_0)/(V_{\rm max}-V_0) \rangle$ do not measure the strength of the
evolution directly \cite{Avni80}, and it would be wrong to use this
test to compare the relative rates of evolution at high and low
redshifts.


\begin{figure}
\psfig{file=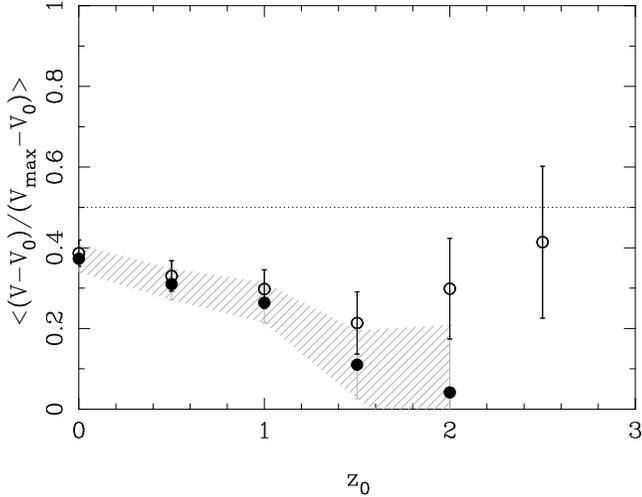,width=\hsize}
\caption{The banded \vvmax\ test for the 2-mJy sample (open circles).
The dotted horizontal line indicates the value of 0.5 expected if
there is no evolution in the comoving density of sources at redshifts
$z>z_0$.  Values of $\langle (V-V_0)/(V_{\rm max}-V_0) \rangle < 0.5$
indicate a deficit of high-redshift objects.  The solid points show
the results of repeating the test after excluding the highest-redshift
source, 53W061, whose photometric redshift is uncertain.  The hatched
region shows the corresponding one-sigma errors.\label{vvmax}}
\end{figure}

In figure~\ref{vvmax} we present the results of applying this test to
the 2-mJy Hercules sample.  Over all redshifts, we find $\langle
(V-V_0)/(V_{\rm max}-V_0) \rangle = 0.40 \pm 0.03$, differing from the
no-evolution line by 3-$\sigma$.  The significance of the result
remains at $\ge 3$-$\sigma$ for all redshifts $z<2$.  At the highest
redshifts, $z>2$, the significance is $\sim 1$-$\sigma$, due to the
small number of sources (5.4 after weighting).  In particular, the
highest-redshift source in the sample (53W061 at $z_{\rm phot}=2.88$)
is the only object at $z>2.5$, and with a weight of 2.35 it has a
disproportionate effect on the test.  We illustrate this in
figure~\ref{vvmax} by removing 53W061 from the sample and repeating
the banded \vvmax\ test.  The value of $\langle (V-V_0)/(V_{\rm
max}-V_0) \rangle$ then falls to $0.04\pm0.17$ at $z>2$.  Recall that
53W061 has two peaks in its photometric redshift likelihood function
and we chose the high-redshift peak on the basis of its identification
as a likely quasar on the Mayall photographic plates.  However, it is
possible that the source is actually at the lower redshift ($z_{\rm
phot}=0.29$); thus excluding it from the \vvmax\ test is not
unreasonable.

The small values of \bvvmax\ at all redshifts point to a genuine
deficit of high-redshift objects in the sample.  It doesn't mean that
there is a decline in the comoving density of sources at low redshift,
but rather a lack of strong positive evolution at low-$z$ in this
population is failing to cancel the high-redshift negative evolution.
For example, in the PSR survey (DP90) there is enough positive
evolution out to $z>1$ to make the initial banded numbers greater than
0.5, until the negative evolution at high redshift begins to dominate
\bvvmax.  In the Hercules sample, however, there is little if any
positive evolution (compare figure~\ref{rlfdata}) and so the high-$z$
deficit dominates \bvvmax\ at all redshifts.  

Could the results of the \vvmax\ test be due to something other than a
high-$z$ decline in the comoving density of sources?  Looking at the
values of \vvmax\ for the individual objects shows that 25\% of the
sample could have been detected at $z>3$, but there are in fact no
sources at such redshifts.  Comparing the open and solid points in
figure~\ref{vvmax} suggests that just a few sources at the highest
redshifts may remove the evidence for evolution at $z>2.5$.  To
investigate this we added to the distribution the three unidentified
sources, placing them at random redshifts ($0.5<z<3$).  This did not
change the result of the test; in fact, placing all three sources at
$z>2$ actually increased the significance of the negative evolution at
the highest redshifts (recall that the error on \bvvmax\ goes as
$N^{-0.5}$).  Therefore the conclusion that there is a significant
decline in the source density cannot be falsely attributed to the
properties of just a few sources.

Another concern is the effect that spectral curvature may have on the
results of the \vvmax\ test.  In calculating $V_{\rm max}$ we assumed
that the 0.6--1.4~GHz spectral indices of the sources remain constant
at higher frequency, but that need not be the case.  If the radio
spectrum steepens at higher frequencies then the luminosity of a
source will be lower than that calculated from the observed spectral
index, and the inferred $V_{\rm max}$ will also be smaller.  This in
turn will give a larger value for \vvmax.  We thus repeated the test
after increasing the spectral index of each source ($\alpha
\rightarrow \alpha + \delta$, where $S_\nu \propto \nu^{-\alpha}$).
For reasonable changes in spectral index, $\delta \simeq 1$, the
results remain essentially unchanged although the significance drops
from 3-$\sigma$ to 2-$\sigma$ at zero redshift.  The spectral indices
must be increased by $\delta \simeq 3$ before the data become
consistent with no evolution at $z<1.5$, and even then values of
\bvvmax~$=0.2$--0.3 are found at higher redshifts.  Thus the results
are robust against spectral curvature.

\section{Conclusions}

Spectroscopic observations of sources in the LBDS Hercules sample have
yielded new redshifts for 11 objects, bringing the redshift content of
the sample to 65\% (47 of 72 sources).  Upper limits to the redshifts
of a further 10 sources have been found, based on the absence of a
redshifted Lyman-limit break in their spectra.

For the remaining one-third of the sample we have estimated their
redshifts from their broadband photometry.  After testing several
estimation methods on our data, we found that the most accurate
results were produced by modelling the spectra as the sum of an old
(14~Gyr) stellar population plus a young (0.1~Gyr) stellar population
that contributes some fraction (\flilly) to the 5000-\AA\ flux.
Comparing the photometric with the spectroscopic redshift for those
sources with a measured redshift showed that our estimates have an
accuracy of $\langle \Delta z \rangle_{\rm med} = 0.13$ and an rms
dispersion of $\sigma_{\rm med} = 0.29$, comparable to that of other
authors \cite{Hogg98}.

We have used the combined spectroscopic plus photometric redshift
distribution of the 2-mJy Hercules sample to investigate the evolution
of the 1.4~GHz radio luminosity function.  The 2-mJy sample, which
excludes the nine sources with $S_{1.4}<2$~mJy and weights $\ga 3$,
has a median redshift of 0.80, comparable to surveys of lower flux
density \cite{Richards98,Windhorst98}.  The fact that the redshift
distribution hardly changes over two orders of magnitude in radio
flux, implies that the faint radio source population must be evolving
fairly strongly -- the fainter surveys are not simply detecting the
same population at a higher redshift.  The low-flux density (sub-mJy,
$\umu$Jy) surveys are dominated by starbursts, whereas the LBDS survey
is dominated by giant elliptical radio galaxies and quasars, at least
at $S_{1.4}\ga 3$~mJy \cite{Kron85,Windhorst98,Hopkins00}.

Dunlop \& Peacock (1990) generated an ensemble of models of the
evolving radio luminosity function of AGN for a sample of sources with
$S_{2.7}\ge 100$~mJy.  The results indicated that the data required a
cut-off in the RLF at $z\ga 2$ for both flat- and steep-spectrum
sources.  The models also predicted the redshift distribution of
sources at a lower flux density limit.  Here we have compared the
redshift distribution of the LBDS Hercules sample with these models,
and found that two of them (FF-4 \& FF-5) are consistent with the new
data.  In both models, the RLF is forced to decline from $z=2$ to zero
at $z=5$, with the form of the cut-off being sinusoidal in FF-5 and
unconstrained (beyond being smooth and continuous) in FF-4.  The other
five models of DP90 are inconsistent with the Hercules sample, in
particular the simple parametric pure luminosity evolution (PLE) and
luminosity/density evolution (LDE) models predicted that the mJy radio
sources would have been at much higher redshifts than the data allow.

The observed RLF was calculated by directly binning the data in
redshift and luminosity, for $z=0$--4 and $\logten P_{2.7}=20$--28.
Comparing the observed RLF with the best-fitting models from DP90,
shows that the low-power ($P_{2.7}<P^*\sim 10^{25}$~\WHzsr) RLF
appears to evolve more slowly than at higher luminosities.  At high
redshifts ($z>1$--2) the observed RLF shows an unambiguous decrease in
the comoving density of sources in all luminosity bins, independent of
the models.

Finally, we investigated the redshift distribution of the 2-mJy
Hercules sample via the model-independent banded \vvmax\ test.  Values
of $\langle (V-V_0)/(V_{\rm max}-V_0) \rangle < 0.5$ at all redshifts
indicate a deficit of high-redshift sources in the sample.  This
result does not change if the three optically-unidentified sources are
actually at high-$z$, nor if the effect of spectral curvature is
considered.

We conclude that the data from the LBDS Hercules sample shows clear
evidence for a decline, or cut-off, in the RLF at high redshifts.  The
low-luminosity population evolves more slowly and may begin to decline
at lower redshifts ($z\sim 1$) than the more powerful sources ($z\sim
2$).  

How do we relate this to the physical evolution of the faint radio
source population?  It is known from synchrotron-aging and energy
arguments that the lifetime of a radio source is of the order
$10^7$--$10^8$ years, thus the RLF does not measure the evolution of
individual sources over cosmological timescales but only that of the
population as a whole.  The evolution in comoving density with
redshift can be explained as a change in the formation rate of radio
sources.  Our results then imply that the low-luminosity ($P_{1.4} \la
10^{26}$~\WHzsr) radio sources: (i) tend to form at a later epoch than
the more powerful sources of DP90; and (ii) continue to form at a
fairly constant rate from $z\sim 1$ until the present day.
Alternatively, the evolution in the RLF can be explained by changes in
the lifetimes of radio sources -- the high-$z$ cut-off could be due to
the radio sources being active for shorter periods at higher
redshifts.

Faint radio galaxies in the 6C survey have been shown to have smaller
half-light radii than the powerful 3CR radio galaxies at $z\sim 1$
(Roche, Eales \& Rawlings 1998)\nocite{Roche98}.  We have similarly
found that two of the sources in the fainter LBDS Hercules sample
(53W069 and 53W091, both at $z\simeq 1.5$) are smaller still, with
half-light radii of $r_e\simeq 2.4$~kpc \cite{Waddington01}.  To the
extent that the size of an elliptical galaxy is a measure of its mass,
this suggests that lower luminosity radio sources form in galaxies of
lower mass.

Combining these results with the evolution that we find in the RLF
allows us to build a picture of radio source activity, if we assume
the evolution is due to number density changes.  The first powerful
radio sources form in massive galaxies at $t\simeq 1$~Gyr ($z\sim 5$)
in our assumed cosmology.  The formation epoch of the first
low-luminosity radio sources is not constrained by our data, but is
presumably no earlier than this (the first quasars also appear at
$t\simeq 1$~Gyr).  The comoving density of massive galaxies that host
powerful radio sources increases rapidly, until it peaks at $t\simeq
2$--3~Gyr ($z\sim 2$) then it falls until the present ($t=13$~Gyr).
The comoving density of low-mass galaxies hosting low-luminosity radio
sources increases until $t\simeq 4$--5~Gyr ($z\sim 1$), then it
appears to remain roughly constant until today.  At any given epoch
the steepness of the RLF means there are always more low-luminosity
sources than high-luminosity ones.  Note that this division into
`high' and `low' luminosity sources is not meant to imply a dual
population -- radio sources have a continuous range of luminosities
and so the redshift of the turnover in the RLF is presumed to also
vary continuously.

This evolution can be understood as follows.  More massive galaxies
have more gas available as fuel for the AGN.  They also have a deeper
potential well, so it is easier for the gas to fall into the centre of
the galaxy.  Therefore it takes less time for sufficient gas to be
accreted to trigger the AGN, and more luminous radio sources
preferentially form at earlier epochs.  With less gas and a shallower
potential well, less-massive galaxies will typically take longer to
become active and will have a lower radio luminosity.

This interpretation of the data is not unique, as density evolution is
not the only possible explanation of the evolving RLF.  For example,
changes in the intergalactic medium with redshift will affect the
propagation of the radio jets which could in turn affect their
lifetimes.  As we noted above, shorter source lifetimes would reduce
the RLF at high-$z$ without needing to invoke number density changes.
However, density evolution of the form outlined above will likely be
an important, if not dominant, factor governing the observed evolution
in the RLF.

\section*{Acknowledgments}

We thank David Koo, Richard Kron, Hyron Spinrad, Arjun Dey and Daniel
Stern for allowing us to use a few unpublished redshifts; and Marek
Kukula for contributing to the WHT observations.  The William Herschel
Telescope is operated on the island of La Palma by the Isaac Newton
Group in the Spanish Observatorio del Roque de los Muchachos of the
Instituto de Astrofisica de Canarias.  This work was supported in part
by a PPARC research studentship to IW; and in part by NSF grants
AST-88211016 \& AST-9802963 and NASA grants GO-5308.01.93A,
GO-5985.01.94A \& GO.7208.01.96A from STScI under NASA contract
NAS5-26555 to RAW.


\label{lastpage}

\bsp

\end{document}